\documentclass[11pt,a4paper]{article}
\usepackage{jcappub}
\usepackage{epsfig}
\usepackage{amsmath}
\usepackage{amssymb}

\def\thebiblio#1{
\begin{center}\bf \large References
\end{center}
\list
{[\arabic{enumi}]}{\settowidth\labelwidth{#1.}\leftmargin\labelwidth
 \advance\leftmargin\labelsep
 \usecounter{enumi}}
 \def\newblock{\hskip .11em plus .33em minus -.07em}
 \sloppy
 \sfcode`\.=1000\relax}

\renewcommand{\[}{\begin{equation}}
\renewcommand{\]}{\end{equation}}
\newcommand{\lsim}{\mbox{\raisebox{-.9ex}{~$\stackrel{\mbox{$<$}}{\sim}$~}}}
\newcommand{\gsim}{\mbox{\raisebox{-.9ex}{~$\stackrel{\mbox{$>$}}{\sim}$~}}}

\title{Inflationary buildup of a vector field condensate and its cosmological consequences}

\author[a]{Juan C. Bueno Sanchez,}
\author[b]{Konstantinos Dimopoulos}%

\affiliation[a]{Departamento de F\'isica At\'omica, Molecular y Nuclear, Universidad Complutense de Madrid,\\
28040 Madrid, Spain}
\affiliation[b]{Consortium for Fundamental Physics, Physics Department, Lancaster University, Lancaster LA1 4YB, UK\\}

\abstract{Light vector fields during inflation obtain a superhorizon
perturbation spectrum when their conformal invariance is appropriately broken.
Such perturbations, by means of some suitable mechanism
(e.g. the vector curvaton mechanism), can contribute to the curvature
perturbation in the Universe and produce characteristic signals, such as
statistical anisotropy, on the microwave sky, most recently surveyed by
the Planck satellite mission. The magnitude of such characteristic features
crucially depends on the magnitude of the vector condensate generated during
inflation. However, the expectation value of this condensate has so-far been
taken as a free parameter, lacking a definite prediction or a physically
motivated estimate. In this paper, we study the stochastic evolution of the
vector condensate and obtain an estimate for its magnitude. Our study is mainly
focused in the supergravity inspired case when the kinetic function and mass of
the vector boson is time-varying during inflation, but other cases are also
explored such as a parity violating axial theory or a non-minimal coupling
between the vector field and gravity. As an example, we apply our findings in
the context of the vector curvaton mechanism and contrast our
results with current observations.}

\begin{document}
\maketitle

\section{Introduction}

The recent cosmological observations of the Planck satellite mission have
largely confirmed the generic predictions of cosmic inflation, even though
they have put substantial tension to and even excluded specific classes of
inflationary models \cite{planck}. Apart from a red spectral index (which was
already known by the WMAP observations) the one significant deviation from the
so-called vanilla predictions of inflation (such as adiabaticity, Gaussianity
and scale invariance) which was found by Planck was statistical
anisotropy in the low-multipoles of the CMB. This suggests that there may be a
preferred direction in space, which is difficult to account for in the
traditional inflationary paradigm, because the latter utilises only scalar
fields, which cannot break isotropy (see however Ref.~\cite{Bartolo:2013msa}).

This is why, in recent years, there is growing interest of the possible role
that vector fields may play in the physics of inflation. Vector fields
naturally break isotropy and are also a necessary ingredient of fundamental
physics and all the theories beyond the standard model \cite{vectoreview}.
However, until recently, their role in inflation has been ignored. The
pioneering work in Ref.~\cite{Dimopoulos:2006ms}
was the first to consider the possible contribution of vector
fields to the curvature perturbation in the Universe. It was soon realised that
such a contribution could be inherently anisotropic and can give rise to
statistical anisotropy \cite{yokosoda} as demonstrated via the
$\delta N$-formalism in Ref.~\cite{Dimopoulos:2008yv}. The degree of
statistical anisotropy due to the direct
contribution of the anisotropic perturbations of a vector field, is crucially
determined by the magnitude of the vector boson condensate, which corresponds
to the homogeneous background zero mode of the vector field. This may generate
indirectly additional statistical anisotropy, by mildly anisotropising the
inflationary expansion leading to anisotropic inflation \cite{anisinf}, which
renders the perturbations of the scalar inflaton field themselves statistically
anisotropic. In this case too, the degree of the anisotropy is determined by
the magnitude of the vector condensate. The latter remains significant and it
is not diluted by inflationary expansion only because it is replenished by
continuous vector field particle production during inflation. The magnitude of
the condensate, however, has been taken as a free parameter in all
considerations so far. This, not only is incomplete and unrealistic but also it
removes constraining power from vector field models, which otherwise can shed
some light on the total duration of inflation, necessary in order to have the
required condensate created. This adds onto the fact that, as mentioned, the
presence of a vector field condensate renders the inflationary expansion mildly
anisotropic, which in turn evades the no-hair theorem and opens potentially a
window to the initial conditions of inflation \cite{anisinf}.

In this paper we develop in detail the techniques necessary to calculate the
stochastic buildup of an Abelian vector field condensate during inflation and
provide specific predictions of the magnitude of such a condensate. We focus
mostly in the case of a time-varying kinetic function and mass, because this
corresponds to a system which is drastically different from the well-known
buildup of a scalar field condensate in Ref.~\cite{BD} and can be also
motivated by supergravity considerations (see for example \cite{sugravec,
Dimopoulos:2009am,jacques}). We apply our findings in the vector curvaton
mechanism of
Ref.~\cite{Dimopoulos:2006ms} (for a review see Ref.~\cite{Dimopoulos:2011ws})
as an example of the predictive power of our results. However, we
also look into other models such as a massless Maxwell vector field with
varying kinetic function (as used in Ref.~\cite{anisinf} for example), an
Abelian vector field non-minimally coupled to gravity through a coupling of the
form $RA^2$ \cite{nonmin} and an axial theory, which also considers the effect
of the $\propto F\tilde F$ term, in the buildup of the vector field condensate
\cite{Dimopoulos:2012av}. At first approximation we consider quasi-de Sitter
inflation, with a subdominant Abelian spectator field.

Throughout our paper, we use natural units with \mbox{$c=\hbar=k_B=1$} and
\mbox{$8\pi G=m_P^{-2}$}, where $G$ is Newton's gravitational constant and
\mbox{$m_P=2.4\times 10^{18}\,$GeV} is the reduced Planck mass.

\section{The importance of a vector field condensate}
\label{sec-stanis}

In this paper we study in detail the stochastic buildup of a vector field
condensate during inflation. The existence of such a condensate may affect
the inflationary expansion and render it mildly anisotropic, thereby evading
the no-hair theorem and generating statistical anisotropy in the inflaton's
perturbations \cite{anisinf}. Moreover, since the buildup of the condensate is
based on the particle production of the vector field perturbations, the
condensate is essential in order to quantify the effect on the curvature
perturbation that the vector perturbations can have directly. To demonstrate
this consider that the statistical anisotropy in the spectrum of the curvature
perturbations can be parametrised as \cite{acw}
\begin{equation}
{\cal P}_\zeta(\mbox{\boldmath $k$})={\cal P}_\zeta^{\rm iso}(k)
[1+g(k)(\mbox{\boldmath $d\cdot\hat k$})^2+\cdots],
\label{acw}
\end{equation}
where ``iso'' denotes the isotropic part, \mbox{\boldmath $d$} is the unit
vector depicting the preferred direction,
\mbox{$\mbox{\boldmath $\hat k$}\equiv\mbox{\boldmath $k$}/k$} is the unit
vector along the wavevector \mbox{\boldmath $k$} (with $k$ being the modulus
of the latter), the ellipsis denotes higher orders
and $g(k)$ is the so-called anisotropy parameter, which quantifies the
statistical anisotropy in ${\cal P}_\zeta$. The latest observations from the
Planck satellite suggest that $g$ can be at most a few percent \cite{planck-}.

For example, if the curvature perturbation is affected by a single scalar and
a single vector field then \cite{Dimopoulos:2008yv}
\begin{equation}
g=\xi\frac{{\cal P}_\|-{\cal P}_+}{{\cal P}_\phi+\xi{\cal P}_+}\,,
\label{g}
\end{equation}
where ${\cal P}_\phi$  and ${\cal P}_\|$ denote the power spectra of the scalar
field $\phi$ (e.g. the inflaton) and the longitudinal component of the vector
field $W_\mu$ respectively, while
\mbox{${\cal P}_+\equiv\frac12({\cal P}_L+{\cal P}_R)$} with ${\cal P}_L$ and
${\cal P}_R$ being the spectra of the left and right polarisations of the
transverse components of the vector field respectively. The parameter $\xi$ is
defined as \mbox{$\xi\equiv N_W^2/N_\phi^2$}, where $N_\phi$ denotes the amount
of modulation of the number of elapsing $e$-foldings because of the scalar
field \mbox{$N_\phi\equiv\partial N/\partial\phi$}, while similarly $N_W$ denotes
the amount of modulation of the number of elapsing $e$-foldings because of the
vector field: \mbox{$N_W=|\mbox{\boldmath $N_W$}|$}, where
\mbox{$N_W^i\equiv\partial N/\partial W_i$}. According to the
$\delta N$-formalism \cite{Dimopoulos:2008yv}, the curvature perturbation is
given by \mbox{$\zeta=N_\phi\delta\phi+N_W^i\delta W_i+\cdots$}, where Einstein
summation over the spatial indices \mbox{$i=1,2,3$} is assumed.
Therefore, the value of $N_W$ is necessary to quantify $g$ (through $\xi$).
This value, in turn, is partly determined by the value of the vector field
condensate, which we investigate in this paper.

For example, in the vector curvaton scenario \cite{Dimopoulos:2006ms}
we have \cite{Dimopoulos:2008yv,Dimopoulos:2011ws}
\begin{equation}
N_W^i=\frac23\hat\Omega_{\rm dec}\frac{W_i}{W^2}\,,
\end{equation}
where \mbox{$\hat\Omega_{\rm dec}=\frac{3\Omega_{\rm dec}}{4-\Omega_{\rm dec}}\sim
\Omega_{\rm dec}$}, with $\Omega_{\rm dec}$ denoting the vector field density
parameter at the time of the
vector field decay. In the above \mbox{$W=|\mbox{\boldmath $W$}|$} is the
magnitude of the vector field condensate and $W_i$ its components.

Similarly, in the end of inflation mechanism, the waterfall at the end of
Hybrid Inflation can be modulated by a vector field \cite{yokosoda}, whose
condensate determines $N_W$. Indeed, in this case \cite{Dimopoulos:2008yv}
\begin{equation}
N_W^i=N_c\frac{\lambda_W}{\lambda_\phi}\frac{W_i}{\phi_c}\,,
\end{equation}
where $\lambda_\phi$ \{$\lambda_W$\} is the coupling of the interaction term
between the waterfall field and the inflaton \{vector\} field
and \mbox{$N_c=\partial N/\partial\phi_c$} with $\phi_c$ being the critical
value of the inflaton when the waterfall occurs. Thus, we see again that
\mbox{$N_W^i\propto W_i$}, i.e. $N_W$ is determined by the magnitude of the
condensate components.

In both the above examples to determine $g$ it is necessary to know $W_\mu$.
The value of the latter until now has been taken as a free parameter. In this
paper we calculate it explicitly by considering the stochastic formation of the
condensate through particle production. Finally, it is important to point out
that, apart from $g$, the components of the condensate also determine the
preferred direction itself, because
\mbox{\mbox{\boldmath $d$}\ =\ \mbox{\boldmath $\hat N_W$}} in Eq.~(\ref{acw})
\cite{Dimopoulos:2008yv}.

\section{Our model}\label{sec_case}
In this section we introduce the vector field model which we want to study. To
illustrate the growth of the vector condensate we consider the
model~\cite{Dimopoulos:2009am} (see also Ref.~\cite{fF2})
\begin{equation}\label{eq67}
{\cal L}=-\frac{1}{4}fF_{\rm \mu\nu}F^{\mu\nu}+\frac{1}{2}m^2A_\mu A^\mu\,,
\end{equation}
where $f$ is the kinetic function, $m$ is the mass of $A_\mu$ and the field
strength tensor is \mbox{$F_{\mu\nu}=\partial_\mu A_\nu - \partial_\nu A_\mu$}.
During inflation, \mbox{$f=f(t)$} and \mbox{$m=m(t)$} can be functions of
cosmic time $t$. Following the convention in Ref.~\cite{Dimopoulos:2009am} we
consider
\[
f\propto a^\alpha\quad,\quad m\propto a^\beta\,.
\]

The motivation for the above model is ample. In supergravity the fundamental
functions of the theory are the scalar and K\"{a}hler potentials and the gauge
kinetic function $f$ of the gauge fields, which is, in principle, a holomorphic
function of the scalar fields of the theory. Now, due to K\"{a}hler corrections,
the scalar fields obtain masses of order the Hubble scale \cite{DRT} so they
are expected to fast-roll down the slopes of the scalar potential leading to a
sizable modulation of $f$. The same is true in the context of superstrict.
Thus, time dependence of the vector field kinetic function is natural to
consider during inflation.\footnote{The above reasons led a plethora of authors
to consider such a model in cosmology, either to generate a primordial magnetic
field \cite{ratra}, or to give rise to anisotropic inflation \cite{anisinf} or
to directly affect the curvature perturbation
\cite{sugravec,Dimopoulos:2009am,yokosoda}.}
Similar considerations also apply for the masses of
vector fields, which can be modulated by varying scalar fields as well. A
D-brane inflation example of this model can be seen in Ref.~\cite{DBIcurv}.

In the context of this paper, though, we will refrain to be grounded on a
specific theoretical background, albeit generic. Instead, we will consider that
\mbox{$f=f(t)$} and \mbox{$m=m(t)$} only, and explore particle production and
the formation of a vector field condensate in its own right.\footnote{Hence, we
will not consider effects due to the coupling of vector and scalar field
perturbations, which may enhance statistical anisotropy \cite{namba}.}
The reason, as we
will show, is that the model demonstrates an untypical behaviour with the
condensate never equilibrating and being dominated by the longitudinal modes,
whose stochastic variation is diminishing with time, in contrast to the scalar
field case, where the variation is $H/2\pi$ per Hubble time and which
equilibrates to the value $\sim H^2/m$ over long enough time \cite{BD}. The
value of
the accumulated condensate is essential in determining observables, such as
statistical anisotropy, in all cases (either when the vector field contributes
to the curvature perturbation directly or indirectly through rendering the
Universe expansion mildly anisotropic).

To study the field dynamics we consider an isotropic inflationary background of
quasi-de Sitter kind, i.e. $H\simeq{\rm cte}$. Assuming $\dot H\simeq 0$, the
equations for the temporal ($A_t$) and spatial (\mbox{\boldmath$A$}) components
of the vector field $A_\mu$ are
\[
\mbox{\boldmath$\nabla$}\cdot\dot{\!\mbox{\boldmath$A$}}-\nabla^2A_t+
\frac{(am)^2}{f}\,A_t=0
\]
and
\[\label{eq54}
\ddot{\!\!\mbox{\boldmath$A$}}+\left(H+\frac{\dot f}{f}\right)
\dot{\!\mbox{\boldmath$A$}}+\frac{m^2}{f}\mbox{\boldmath$A$}-a^{-2}
\nabla^2\mbox{\boldmath$A$}
=\left(\frac{\dot f}{f}-2\frac{\dot m}{m}-2H\right)\nabla A_t\,.
\]

Since inflation homogenises the vector field $A_\mu$, we impose the condition
$\partial_iA_\mu=0$, which then translates into
$A_t=0$~\cite{Dimopoulos:2006ms}. Nevertheless, particle production during
inflation gives rise to perturbations of the vector field
$\delta A_\mu\equiv(\delta A_t,\delta\mbox{\boldmath$A$})$
\[\label{eq52}
A_\mu(t,\mbox{\boldmath$x$})=A_\mu(t)+\delta A_\mu(t,\mbox{\boldmath$x$})\,,
\]
which we expand in Fourier modes
$\delta {\cal A}_\mu\equiv(\delta {\cal A}_t,\delta\mbox{\boldmath${\cal A}$})$
as
\[\label{eq23}
\delta A_\mu(t,\mbox{\boldmath$x$})=
\int\frac{d^3\mbox{\boldmath$k$}}{(2\pi)^{3/2}}\,
\delta{\cal A}_\mu(t,\mbox{\boldmath$k$})\,\exp(i\mbox{\boldmath$k\cdot x$})
\,.
\]
Because $A_t=0$ for the background vector field, the temporal component is
itself a perturbation, i.e.
$A_t(t,\mbox{\boldmath$x$})=\delta A_t(t,\mbox{\boldmath$x$})$, determined by
the spatial field perturbations
\[\label{eq16}
\delta{\cal A}_t+\frac{i\partial_t\left(\mbox{\boldmath$k$}\cdot
\delta\mbox{\boldmath${\cal A}$}\right)}{k^2+(am)^2/f}=0\,.
\]
At this point we introduce the physical vector field
\[\label{eq53}
\mbox{\boldmath$W$}\equiv\sqrt f\mbox{\boldmath$A$}/a\,.
\]
Writing Eq.~(\ref{eq54}) in terms of the physical vector field perturbation
$\delta\mbox{\boldmath$W$}$ we have
\[\label{eq19}
\delta\mbox{\boldmath$\ddot{W}$}+3H\delta
\mbox{\boldmath$\dot{W}$}+\left[-\frac14(\alpha+4)
(\alpha-2)H^2+M^2-a^{-2}\nabla^2\right]\,\delta
\mbox{\boldmath$W$}=p(t)\mbox{\boldmath $\nabla$}A_t\,,
\]
where
\[
p(t)=\left(\frac{\dot f}{f}-\frac{2\dot m}{m}-2H\right)\sqrt{f}\,a^{-1}
\]
and
\[\label{eq79}
M\equiv m/\sqrt{f}\propto a^{\beta-\alpha/2}
\]
is the time-dependent effective mass of the physical vector field.

To study the quantum production of the vector field we introduce
creation/annihilation operators for each polarisation as follows
\[\label{eq68}
\delta\mbox{\boldmath $W$}(t,\mbox{\boldmath$x$})=\sum_{\lambda=L,R,\parallel}
\delta\mbox{\boldmath $W$}_\lambda(t,\mbox{\boldmath$x$})\,,
\]
where
\begin{equation}
\label{eq4}
\delta\mbox{\boldmath $W$}_\lambda(t,\mbox{\boldmath$x$})\equiv
\int\frac{d^3k}{(2\pi)^3}\left[
\mbox{\boldmath $e$}_\lambda(\hat{\mbox{\boldmath $k$}})
\hat a_\lambda(\mbox{\boldmath $k$})
w_\lambda(t,k)e^{i\mbox{\scriptsize\boldmath$k$}\cdot \mbox{\scriptsize
\boldmath$x$}}+\mbox{\boldmath $e$}^*_\lambda(\hat{\mbox{\boldmath $k$}})
\hat a^\dag_\lambda(\mbox{\boldmath $k$})
w^*_\lambda(t,k)e^{-i\mbox{\scriptsize\boldmath$k$}\cdot \mbox{\scriptsize
\boldmath$x$}}
\right]\,,
\end{equation}
and where $\hat{\mbox{\boldmath $k$}}\equiv\mbox{\boldmath $k$}/k$,
$k\equiv|\mbox{\boldmath $k$}|$ and $\lambda=L,R,\parallel$ labels the
{\it Left} and {\it Right} transverse and longitudinal polarisations
respectively. The polarisation vectors are
\begin{equation}\label{eq30}
\mbox{\boldmath$e$}_L\equiv\frac{1}{\sqrt2}(1,i,0),\;\;
\mbox{\boldmath$e$}_R=\frac{1}{\sqrt2}(1,-i,0),\;\;
\mbox{\boldmath$e$}_\parallel=(0,0,1)\,.
\end{equation}
The perturbation $\delta\mbox{\boldmath$W$}(t,\mbox{\boldmath$x$})$ is
quantised imposing equal-time commutation relations
\[\label{eq29}
[\hat a_\alpha(\mbox{\boldmath $k$}),
\hat a^\dag_{\beta}(\mbox{\boldmath $k$}')]=(2\pi)^3
\delta(\mbox{\boldmath $k$}-\mbox{\boldmath $k$}')\delta_{\alpha\beta}\,,
\]
whereby quantum particle production is uncorrelated among different
polarisation modes, i.e.
$\langle\delta\mbox{\boldmath$W$}_\alpha(t,\mbox{\boldmath$x$})\,
\delta\mbox{\boldmath$W$}_\beta(t',\mbox{\boldmath$y$})
\rangle\propto\delta_{\alpha\beta}$.

\subsection{Case $f\propto a^{-4}$ and $m\propto a$}\label{sec1}

The reason to focus our attention in this case is twofold. Firstly, the vector
field obtains a nearly scale-invariant spectrum of superhorizon perturbations.
This means that its effects, e.g. by generating statistical anisotropy in the
curvature perturbation, are apparent (and the same) in all scales. Also, there
is no ``special time'' during inflation (i.e. no fine-tuning), when particle
production is more pronounced. Thus, the only relevant variable is the total
inflationary $e$-foldings. Apart from simplicity, however, it has been shown
that the above behaviour can be an attractor solution if $f$ and $m$ are
modulated by the rolling inflaton field, because vector backreaction can adjust
the variation of the inflaton accordingly \cite{jacques}.
%

The second reason to consider such a choice is that it constitutes substantial
deviation with respect to the case of a minimally coupled, light scalar
field~\cite{Starobinsky:1982ee}. As we show later on, and in contrast to the
case of a light scalar field, the vector field features a non-trivial
superhorizon evolution. Moreover, the longitudinal and transverse modes of the
vector field obtain different superhorizon perturbation spectra, which then
must be treated separately.

Introducing the expansion (\ref{eq4}) into Eq.~(\ref{eq19}) and taking into
account that $\alpha=-4$ we obtain the evolution equations for the transverse
and longitudinal mode functions \cite{Dimopoulos:2009am}
\begin{equation}\label{eq15}
\ddot w_{L,R}+3H\dot w_{L,R}+
\left(\frac{k^2}{a^2}+M^2\right)w_{L,R}=0\,,
\end{equation}
\begin{equation}\label{eq8}
\ddot w_\parallel+
\left(3+\frac{8}{1+r^2}\right)H\dot w_\parallel+
\left[\frac{24}{1+r^2}H^2+\left(\frac{k}{a}\right)^2(1+r^2)\right]
w_\parallel = 0\,.
\end{equation}
where $r\equiv\frac{aM}{k}$. In the limit $r\gg r_c\gg1$, where $r_c$ is
defined for a given $k$ by the condition that the terms in the square brackets
in Eq.~(\ref{eq8}) become comparable \cite{Dimopoulos:2009am}, the equations
for $w_{L,R}$ and $w_\parallel$ coincide. The solutions to the above equations in
the aforementioned limit are found to be
\[\label{eq17}
w_{L,R}(t,k)=a^{-3/2}\left[\widehat c_1J_{1/2}\left(\frac{M}{3H}\right)+
\widehat c_2J_{-1/2}\left(\frac{M}{3H}\right)\right]
\]
\[\label{eq13}
w_\parallel(t,k)=a^{-3/2}\left[\widehat c_3J_{1/2}\left(\frac{M}{3H}\right)+
\widehat c_4J_{-1/2}\left(\frac{M}{3H}\right)\right]\,,
\]
where the constants $\widehat c_i$ are determined by
\[
\widehat c_1=\frac{i}{2}\sqrt{\frac{\pi}{H}}
\left(\frac{aH}{k}\right)^{3/2}\left(\frac{3H}{M}\right)^{1/2}
\quad,\quad \widehat c_2=\frac16\sqrt{\frac{\pi}{H}}
\left(\frac{k}{aH}\right)^{3/2}\left(\frac{M}{3H}\right)^{1/2}\,,
\]
\[\label{eq14}
\widehat c_3=i\, \widehat c_2
\quad\textrm,\quad \widehat c_4=i\, \widehat c_1\,.
\]

In view of Eq.~(\ref{eq15}) the transverse modes $w_{L,R}$ behave like a
minimally coupled, massive scalar field. Therefore, provided $M\ll H$, the
transverse modes $w_{L,R}$ cease to oscillate on superhorizon scales
($k/aH\ll1$) and obtain an expectation value
\[\label{eq24}
w_{L,R}\simeq \frac{i}{\sqrt{2k}}\left(\frac{H}{k}\right)\,.
\]
Also, the first and second derivatives give
\[\label{eq55}
\dot{w}_{L,R}\simeq-\frac{M^2}{9H}\,w_{L,R}\quad,
\quad\ddot{w}_{L,R}\simeq -\frac{2M^2}{3}\,w_{L,R}\,.
\]

Regarding Eq.~(\ref{eq8}), although this coincides with Eq.~(\ref{eq15}) in the
limit \mbox{$r\gg r_c\gg1$} (as previously noticed), the longitudinal mode
function $w_\parallel$ does not feature the same superhorizon evolution as
$w_{L,R}$ due to the different boundary conditions imposed in the subhorizon
limit $k/aH\to\infty$ \cite{Dimopoulos:2009am}. Owing to this, and in contrast
to $w_{L,R}$ (determined by the growing mode $\propto J_{1/2}(M/3H)$), the
superhorizon evolution of $w_\parallel$ is dominated by the decaying mode
$\propto J_{-1/2}(M/3H)$. In the limit $r\gg r_c\gg1$ we find
\[\label{eq11}
w_\parallel\simeq i\,w_{L,R}\left(\frac{3H}{M}\right)\,.
\]
Since particle production demands $M\ll H$, the above implies
$|w_\parallel|\gg|w_{L,R}|$, and the vector field is approximately curl-free on
superhorizon scales. Moreover, since $M(t)\propto a^3$ the longitudinal modes
feature a fast-roll evolution on superhorizon scales. Owing to this non-trivial
evolution we find
\[\label{eq5}
\dot{w}_\parallel\simeq -3Hw_\parallel\quad,
\quad\ddot{w}_\parallel\simeq 9H^2w_\parallel\,.
\]

A similar result arises in the case of a heavy scalar field. If we consider a
scalar field $\phi$ with mass \mbox{$m_\phi\gg H$}, the amplitude of the mode
$\phi_k$ varies as $a^{-3/2}$ on superhorizon scales. In fact, such a scaling
begins when the mode is still subhorizon. Consequently, one finds
$2\dot\phi_k\simeq -3H\phi_k$ and $4\ddot\phi_k\simeq 9H^2\phi_k$, similarly to
Eq.~(\ref{eq5}). However, a heavy scalar field does not become classical on
superhorizon
scales~\cite{Lyth:1984gv,Guth:1985ya,Mijic:1994vv,Mijic:1998if,Casini:1998wr}.
In our case, though, the vector field (which remains light) indeed becomes
classical because the occupation number of the $k$-modes grows larger than
unity. Moreover, owing to the factor $3H/M\gg1$ in Eq.~(\ref{eq11}), the
occupation number for longitudinal modes is much larger than the corresponding
to transverse modes.

\section{Stochastic formalism}\label{sec-stocha}
The stochastic approach to inflation
\cite{Starobinsky:1982ee,Vilenkin:1982wt,Linde:1982uu} describes the evolution
of the inflaton field on patches of superhorizon size during inflation from a
probabilistic point of view. The probabilistic nature of the field's evolution
on superhorizon scales owes to the quantum particle production undergone by the
inflaton field during inflation. Quantum particle production, however, is not
exclusive of the inflaton field, but it can be undergone by any light field
during inflation as long as it is not conformally coupled to
gravity~\cite{Birrell:1982ix}. Consequently, the stochastic approach to the
inflaton can be extended to fields other than the inflaton, even if such fields
are subdominant during inflation. This is the case we consider in this paper:
the vector field $A_\mu$ remains subdominant during inflation.\footnote{We do
not consider here the case of vector inflation, which assumes that inflation is
driven by hundreds of randomly oriented vector fields \cite{vecinf}. Neither do
we consider gauge-flation \cite{gaugeflation}. However, our results are readily
extendable in these cases, assuming that there is a quasi-de Sitter background.}

The essence of the stochastic approach to inflation consists in establishing a
divide to separate the long and short distance behavior of the field. Such a
long/short wavelength decomposition is carried out by introducing a time
dependent cut-off scale $k_s\equiv\epsilon a(t) H$, where $\epsilon$ is an
auxiliary parameter that determines the scale at which the separation is
performed. Such scale is usually referred to as the \textit{smoothing} or
\textit{coarse-graining} scale. In the simplest approach, which we follow here,
the long ($k\ll k_s$) and short ($k\gg k_s$) wavelength parts of the field are
split up through a top-hat window function, which implies a sharp transition
between the two regimes\footnote{Other window functions, when applied to
separate the long/short distance behavior of the inflaton field, have been
shown to modify the CMB angular power spectrum at low multipoles
\cite{Liguori:2004fa}.}. Following this approach we decompose the physical
vector field $\mbox{\boldmath$W$}(t,\mbox{\boldmath$x$})$ as follows
\begin{eqnarray}
\mbox{\boldmath $W$}(t,\mbox{\boldmath $x$})&=&
\mbox{\boldmath $W$}_c(t,\mbox{\boldmath $x$})+
\mbox{\boldmath $W$}_q(t,\mbox{\boldmath $x$})\,,\label{eq10}\\
\mbox{\boldmath $W$}_q(t, \mbox{\boldmath $x$})&=&
\int\frac{d^3k}{(2\pi)^
3}\;\theta(k-k_s)\sum_\lambda\left[
\mbox{\boldmath $e$}_\lambda(\hat{\mbox{\boldmath $k$}})
\hat a_\lambda(\mbox{\boldmath $k$})
w_\lambda e^{i\mbox{\scriptsize\boldmath$k$}\cdot \mbox{\scriptsize
\boldmath$x$}}+\mbox{\boldmath $e$}_\lambda^*(\hat{\mbox{\boldmath $k$}})
\hat a_\lambda^\dagger(\mbox{\boldmath $k$})
w_\lambda^* e^{-i\mbox{\scriptsize\boldmath$k$}\cdot \mbox{\scriptsize
\boldmath$x$}}\right]\,,\label{eq3}
\end{eqnarray}
where $\mbox{\boldmath $W$}_c$ \{$\mbox{\boldmath $W$}_q$\} represents the long
\{short\} wavelength part of the field. Although $\mbox{\boldmath $W$}_c$ is
only approximately homogeneous, for it contains modes with
\mbox{$0\leq k\leq k_s(t)$}, according to the separate Universe approach
\cite{Wands:2000dp} and for the sake of simplicity we disregard its spatial
dependence and consider it homogeneous in patches of superhorizon size.
Introducing the decomposition (\ref{eq10}) into Eq.~(\ref{eq54}) we arrive at
the effective equation of motion for $\mbox{\boldmath$W$}_c$
\begin{equation}\label{eq12}
\mbox{\boldmath$\ddot W$}_c+3H\mbox{\boldmath$\dot W$}_c+
\Big(M^2-a^{-2}\nabla^2\Big)\,
\mbox{\boldmath$W$}_c-p(t)\mbox{\boldmath $\nabla$}(A_t)_c=
\mbox{\boldmath $\xi$}(t,\mbox{\boldmath $x$})\,,
\end{equation}
where the source term $\mbox{\boldmath $\xi$}(t,\mbox{\boldmath $x$})$ encodes
the behavior of short-wavelength modes and is determined by
\[
\mbox{\boldmath $\xi$}(t,\mbox{\boldmath $x$})=
-\left\{\mbox{\boldmath$\ddot W$}_q+3H\mbox{\boldmath$\dot W$}_q+
\Big[M(t)-a^{-2}\nabla^2\Big]\,
\mbox{\boldmath$W$}_q-p(t)\mbox{\boldmath $\nabla$}(A_t)_q\right\}\,.
\]
In turn, this can be expressed as the superposition of a number of sources (one
per polarisation mode) such that $\mbox{\boldmath $\xi$}(t,\mbox{\boldmath$x$})
=\sum_\lambda\mbox{\boldmath $\xi$}_\lambda$, where
\begin{eqnarray}\label{eq40}
\mbox{\boldmath $\xi$}_\lambda(t,\mbox{\boldmath$x$})&\equiv&
-\int\frac{d^3k}{(2\pi)^
3}\Bigg\{\left[\ddot{\theta}(k-k_s)+3H\dot{\theta}(k-k_s)\right]
\left[\mbox{\boldmath $e$}_\lambda
\hat a_\lambda(\mbox{\boldmath $k$})w_\lambda
e^{i\mbox{\scriptsize\boldmath$k$}\cdot \mbox{\scriptsize
\boldmath$x$}}+\mbox{\boldmath $e$}_\lambda^*\hat a_\lambda^\dagger
(\mbox{\boldmath $k$})w_\lambda^*
e^{-i\mbox{\scriptsize\boldmath$k$}\cdot \mbox{\scriptsize
\boldmath$x$}}\right]\nonumber
\\&&+\,2\dot{\theta}(k-k_s)
\left[\mbox{\boldmath $e$}_\lambda
\,\hat a_\lambda(\mbox{\boldmath $k$})
\dot{w}_\lambda e^{i\mbox{\scriptsize\boldmath$k$}\cdot \mbox{\scriptsize
\boldmath$x$}}+
\mbox{\boldmath $e$}_\lambda^*\hat a_\lambda^\dagger(\mbox{\boldmath $k$})
\dot w_\lambda^* e^{-i\mbox{\scriptsize\boldmath$k$}\cdot \mbox{\scriptsize
\boldmath$x$}}\right]\Bigg\}\,.
\end{eqnarray}

As already mentioned, the probabilistic nature of the field's evolution stems
from the quantum production of field perturbations, which in turn originate
from the field's vacuum fluctuation.\footnote{We do not consider here that the
field perturbations correspond to amplifications of excited states, as e.g. in
Ref.~\cite{amjad}.} Since the probability distribution of the
vacuum fluctuation is gaussian with zero mean, the field's probabilistic
evolution can be accounted for by considering a stochastic source of white
noise with zero mean, i.e. $\mbox{\boldmath $\xi$}(t,\mbox{\boldmath$x$})$ is
such that
\[
\langle\mbox{\boldmath $\xi$}(t,\mbox{\boldmath$x$})\rangle=0\quad,
\quad\langle\mbox{\boldmath $\xi$}(t,\mbox{\boldmath$x$})\,
\mbox{\boldmath $\xi$}(t',\mbox{\boldmath$y$})\rangle\propto
\delta(\mbox{\boldmath$x$}-\mbox{\boldmath$y$})\,\delta(t-t')\,.
\]

Since we are following the separate Universe approach, we are entitled to
neglect the gradient term $a^{-2}\nabla^2\mbox{\boldmath$W$}_c$ in
Eq.~(\ref{eq12}), which is the usual strategy when dealing with scalar
fields. Nevertheless, in our case another gradient term appears in the
evolution equation: $\mbox{\boldmath $\nabla$}(A_t)_c$. Although it is
reasonable to expect that the term in $\mbox{\boldmath $\nabla$}(A_t)_c$ can be
neglected as well in Eq.~(\ref{eq12}), it is instructive to compute such a term
explicitly and compare it with the rest of the terms in (\ref{eq12}). We
perform this in Appendix~\ref{appA}.

After neglecting the gradient terms, the approximate equation of motion for the
coarse-grained vector field $\mbox{\boldmath$W$}_c$ is
\begin{equation}\label{eq28}
\mbox{\boldmath$\ddot W$}_c+3H\mbox{\boldmath$\dot W$}_c+M^2\,
\mbox{\boldmath$W$}_c=\mbox{\boldmath $\xi$}(t,\mbox{\boldmath $x$})\,.
\end{equation}

Although this equation is formally the same as that of a coarse-grained massive
scalar field, the evolution for the vector field requires careful attention
given the existence of polarisation modes and the different perturbation
spectra and superhorizon evolutions. In the next section we explain how to
circumvent such a difficulty and study the stochastic field evolution for the
different polarisation modes in a unified manner.

\subsection{Effective evolution equations}\label{sec-eee}
Since different polarisation modes obey different equations, it is convenient
to separate their contribution to the coarse-grained field. We then introduce
the $\lambda$-polarised coarse-grained vector field
\mbox{\boldmath$W$}$_\lambda$ as follows
\begin{equation}\label{eq51}
\mbox{\boldmath$W$}_\lambda\equiv\int\frac{d^3k}{(2\pi)^
3}\;\theta(k_s-k)\left[\mbox{\boldmath$e$}_\lambda(\hat{\mbox{\boldmath $k$}})
\hat a_\lambda(\mbox{\boldmath $k$})
w_\lambda(t,k)e^{i\mbox{\scriptsize\boldmath$k$}\cdot \mbox{\scriptsize
\boldmath$x$}}+\mbox{\boldmath$e$}_\lambda^*(\hat{\mbox{\boldmath $k$}})
\hat a_\lambda^\dagger(\mbox{\boldmath $k$})w_\lambda^*(t,k)
e^{-i\mbox{\scriptsize\boldmath$k$}\cdot \mbox{\scriptsize
\boldmath$x$}}\right]\,,
\end{equation}
such that $\mbox{\boldmath$W$}_c=\sum_\lambda\mbox{\boldmath$W$}_\lambda$. Owing
to the linearity of Eq.~(\ref{eq28}) we obtain a decoupled system of equations,
one for each polarisation
\begin{equation}\label{eq1}
\mbox{\boldmath$\ddot W$}_\lambda+3H\mbox{\boldmath$\dot W$}_\lambda+M^2
\mbox{\boldmath$W$}_\lambda=\mbox{\boldmath $\xi$}_\lambda\,.
\end{equation}

At this point it is important to recall that, owing to the different boundary conditions obeyed by the various polarisation modes $w_\lambda$, the growing \{decaying\} part of the longitudinal modes ($w_\parallel$) behaves as the decaying \{growing\} part of the transverse modes ($w_\perp$) on superhorizon scales [c.f. Eqs.~(\ref{eq15}) and (\ref{eq8})]. Therefore, on superhorizon scales the growing \{decaying\} mode dominates the superhorizon evolution of the transverse \{longitudinal\} part of the field. This reversal of roles between the longitudinal and transverse modes on superhorizon scales renders Eq.~(\ref{eq1}) inappropriate to describe the stochastic evolution of the longitudinal vector $\mbox{\boldmath$W$}_\parallel$. This is an important difficulty since the evolution of the classical vector field $\mbox{\boldmath$W$}_c$ is dominated the longitudinal part $\mbox{\boldmath$W$}_\parallel$. The reason behind this failure is that the growing part of the solution to Eq.~(\ref{eq1}) (for the longitudinal component) is sourced by $\mbox{\boldmath$\xi$}_\lambda$, which, in turn, is determined by the decaying mode. Being constant on superhorizon scales, the growing mode soon dominates the evolution of $\mbox{\boldmath$W$}_\lambda$, thus leading to an incorrect evolution. In summary, encoding the short-distance behaviour of a massive vector field by means of a stochastic noise source characterised by its two-point function only results in a loss of information, concerning the boundary conditions imposed on the various polarisation modes in the subhorizon regime, that is crucial to properly describe the evolution of the classical field $\mbox{\boldmath$W$}_c$. Of course, one can always find the particular solution to Eq.~(\ref{eq1}) and remove the growing mode by hand, which solves the problem in a rather ad hoc manner.

Apart from the above, and as anticipated at the end of Sec.~\ref{sec1}, there exists yet another complication related to the left-hand side of Eq.~(\ref{eq1}). The stochastic growth of fields during inflation proceeds due to quantum particle production, which in turn demands that $M\ll H$. In the scalar field case, particle production thus implies a slow-roll motion that allows us to neglect second time derivatives on superhorizon scales. Nevertheless, in the case of a massive vector field, when the longitudinal component is physical, one cannot afford such a carelessness. The reason is that \mbox{\boldmath$\ddot W$}$_\lambda$ results in a term of order $M^2\mbox{\boldmath$W$}_\lambda$ for the transverse components and of order $H^2\mbox{\boldmath$W$}_\lambda$ for the longitudinal one. Using Eqs.~(\ref{eq55}) and (\ref{eq5}) as guidance, a rough estimate suggests that $\mbox{\boldmath$\ddot W$}_{L,R}\sim-\frac{2M^2}{3}\mbox{\boldmath$W$}_{L,R}$ and
$\mbox{\boldmath$\ddot W$}_\parallel\sim9H^2\mbox{\boldmath$W$}_\parallel$. Consequently, $\mbox{\boldmath$\ddot W$}_\parallel$ cannot be absentmindedly thrown away even if the vector field is light enough to be produced during inflation. Despite this shortcoming, one might still insist in using Eq.~(\ref{eq1}) as a starting point for the stochastic analysis. The basis to stick to this point of view relies on the fact that, on sufficiently superhorizon scales, the evolution equation of transverse and longitudinal modes coincides. Therefore, consistency demands that the second order equation for the various polarisation modes $\mbox{\boldmath$W$}_\lambda$ be the same.

Our purpose now is to develop the stochastic formalism for vector fields able to address the aforementioned difficulties while using the same second order equation for $\mbox{\boldmath$W$}_\lambda$ as a starting point. The approach followed below consists in introducing the coarse-grained conjugate momentum $\mbox{\boldmath$\Pi$}_\lambda$, thus reducing the second order equation to a system of first order equations, and then eliminating $\mbox{\boldmath$\Pi$}_\lambda$ utilising the superhorizon behaviour of the perturbation modes $w_\lambda$. Following this procedure we manage to arrive at a \emph{single} first order equation providing a correct description of the stochastic evolution of $\mbox{\boldmath$W$}_\lambda$. We want to emphasise that our method goes beyond the Hamiltonian description of stochastic inflation ~\cite{Sasaki:1987gy}. Following the latter, one arrives at first order system leading to a Fokker-Planck equation in the variables $\mbox{\boldmath$W$}_\lambda$ and $\mbox{\boldmath$\Pi$}_\lambda$. In our case, however, we manage to obtain a \emph{single} first order equation for $\mbox{\boldmath$W$}_\lambda$ leading to a Fokker-Planck equation in the variable $\mbox{\boldmath$W$}_\lambda$ only. Apart from this, our procedure can be successfully applied to scalar fields with a non-negligible scale-dependence (i.e. the case of a heavy field) and also away from the slow-roll regime when second-time derivatives cannot be neglected \cite{Juan}. In the following we provide the details of our method.

As advertised, our approach towards a \emph{single} first order equation for
$\mbox{\boldmath$W$}_\lambda$ consists in introducing the coarse-grained conjugate momentum
\begin{equation}\label{eq70}
\mbox{\boldmath$\Pi$}_\lambda\equiv\int\frac{d^3k}{(2\pi)^
3}\;\theta(k_s-k)\left[\mbox{\boldmath$e$}_\lambda(\hat{\mbox{\boldmath $k$}})
\hat a_\lambda(\mbox{\boldmath $k$})
\dot w_\lambda(t,k)e^{i\mbox{\scriptsize\boldmath$k$}\cdot \mbox{\scriptsize
\boldmath$x$}}+\mbox{\boldmath$e$}_\lambda^*(\hat{\mbox{\boldmath $k$}})
\hat a_\lambda^\dagger(\mbox{\boldmath $k$})\dot w_\lambda^*(t,k)
e^{-i\mbox{\scriptsize\boldmath$k$}\cdot \mbox{\scriptsize
\boldmath$x$}}\right]\,.
\end{equation}
After neglecting gradient terms, the equivalent first order stochastic equations are\footnote{In the context of
scalar fields, separate stochastic equations for the field and its conjugate
momentum were introduced in Ref.~\cite{Sasaki:1987gy}. }
\[\label{eq81}
\mbox{\boldmath$\dot \Pi$}_\lambda+3H\mbox{\boldmath$\Pi$}_\lambda+M^2
\mbox{\boldmath$W$}_\lambda=\mbox{\boldmath $\xi$}_{\pi_\lambda}\,,
\]
\[\label{eq82}
\mbox{\boldmath$\dot W$}_\lambda=\mbox{\boldmath$\Pi$}_\lambda+\mbox{\boldmath $\xi$}_{W_\lambda}
\]
where
\[
\mbox{\boldmath $\xi$}_{\pi_\lambda}=-\int\frac{d^3k}{(2\pi)^3}\,\dot\theta(k-k_s)
\left[\mbox{\boldmath $e$}_\lambda(\hat{\mbox{\boldmath $k$}})
\hat a_\lambda(\mbox{\boldmath $k$})
\dot w_\lambda e^{i\mbox{\scriptsize\boldmath$k$}\cdot \mbox{\scriptsize
\boldmath$x$}}+\mbox{\boldmath $e$}^*_\lambda(\hat{\mbox{\boldmath $k$}})
\hat a^\dag_\lambda(\mbox{\boldmath $k$})
\dot w^*_\lambda e^{-i\mbox{\scriptsize\boldmath$k$}\cdot \mbox{\scriptsize
\boldmath$x$}}\right]\,,
\]
\[\label{eq83}
\mbox{\boldmath $\xi$}_{W_\lambda}=-\int\frac{d^3k}{(2\pi)^3}\,\dot\theta(k-k_s)
\left[\mbox{\boldmath $e$}_\lambda(\hat{\mbox{\boldmath $k$}})
\hat a_\lambda(\mbox{\boldmath $k$})
w_\lambda e^{i\mbox{\scriptsize\boldmath$k$}\cdot \mbox{\scriptsize
\boldmath$x$}}+\mbox{\boldmath $e$}^*_\lambda(\hat{\mbox{\boldmath $k$}})
\hat a^\dag_\lambda(\mbox{\boldmath $k$})
w^*_\lambda e^{-i\mbox{\scriptsize\boldmath$k$}\cdot \mbox{\scriptsize
\boldmath$x$}}\right]
\]
are stochastic noise sources for $\mbox{\boldmath$\Pi$}_\lambda$ and
$\mbox{\boldmath$W$}_\lambda$, respectively. Comparing with Eq.~(\ref{eq40}) it is straightforward to show that $\mbox{\boldmath$\xi$}_\lambda=a^{-3}(a^3\mbox{\boldmath$\xi$}_{W_\lambda})^\cdot+\mbox{\boldmath$\xi$}_{\pi_\lambda}$.

Instead of deriving a Fokker-Planck equation for the probability density (as a function of $\mbox{\boldmath$W$}_\lambda$ and $\mbox{\boldmath$\Pi$}_\lambda$) from the first order system (\ref{eq81})-(\ref{eq82}), our approach consists in using the solution to the mode functions $w_\lambda$ to reduce the
first order  system. Indeed, solving the equation of motion for $w_\lambda$ amounts to solving Eq.~(\ref{eq81}), whereas Eq.~(\ref{eq82}) becomes an identity. Nevertheless, we proceed to manipulate Eq.~(\ref{eq82}) to eliminate $\mbox{\boldmath$\Pi$}_\lambda$, thus arriving at a first order equation in $\mbox{\boldmath$W$}_\lambda$. The essence of our method boils down to utilize the superhorizon behavior of $w_\lambda$ to find the function
\[
{\cal F}_\lambda\equiv-\lim_{k/aH\to0}\frac{\dot w_\lambda}{w_\lambda}\,,
\]
in general time-dependent, that allows us to write $\mbox{\boldmath$\Pi$}_\lambda$ in terms of \mbox{\boldmath$W$}$_\lambda$. Consequently,
Eq.~(\ref{eq82}) can be rewritten as
\[\label{eq32}
\mbox{\boldmath$\dot W$}_\lambda+{\cal F}_\lambda(t)\,
\mbox{\boldmath$W$}_\lambda\simeq\mbox{\boldmath $\xi$}_{W_\lambda}\,,
\]
thus dispatching non-negligible second derivatives in Eq.~(\ref{eq1}) and paving the road towards a Fokker-Planck equation for the probability density as a function of $\mbox{\boldmath$W$}_\lambda$ only. Of course, we could proceed similarly eliminating $\mbox{\boldmath$W$}_\lambda$ from Eq.~(\ref{eq81}) to obtain
\[
\mbox{\boldmath$\dot\Pi$}_\lambda+\left(3H-M^2{\cal F}_\lambda^{-1}\right)\mbox{\boldmath$\Pi$}_\lambda\simeq\mbox{\boldmath$\xi$}_{\pi_\lambda}\,,
\]
and derive a Fokker-Planck equation for $\rho(\mbox{\boldmath$\Pi$}_\lambda)$. However, in the following we are simply concerned with $\mbox{\boldmath$W$}_\lambda$, and hence we will work with Eq.~(\ref{eq32}) only.

For the transverse components, since Eq.~(\ref{eq55}) implies
$\mbox{\boldmath$\Pi$}_{L,R}^{(s)}\simeq-\frac{M^2}{9H}\mbox{\boldmath$W$}_{L,R}$
we arrive at ${\cal F}_{L,R}(t)\equiv \frac{M^2}{9H}$, whereas for the
longitudinal component Eq.~(\ref{eq5}) implies
$\mbox{\boldmath$\Pi$}_\parallel^{(s)}\simeq-3H\mbox{\boldmath$W$}_\parallel$ and we
obtain ${\cal F}_\parallel\equiv 3H$.

Regarding the stochastic source, using Eq.~(\ref{eq83}) and the commutation
relations in Eq.~(\ref{eq29}) the self-correlation function can be readily
computed to be
\[\label{eq84}
\langle\mbox{\boldmath$\xi$}_{W_\alpha}(t)\,\mbox{\boldmath$\xi$}_{W_\beta}(t')
\rangle={\cal D}_\alpha\,\delta_{\alpha\beta}\,\delta(t-t')
\]
where
\[\label{eq85}
{\cal D}_\alpha\equiv H\left(\lim_{k\to k_s}\frac{1}{2\pi^2}k^3|w_\alpha|^2\right)
\]
is the diffusion coefficient of the $\lambda$-polarised vector. The parenthesis
represents the perturbation spectrum of $\mbox{\boldmath$W$}_\lambda$ at the
coarse-graining scale, and coincides with the spectrum at horizon crossing when
perturbation spectrum is flat. Using the superhorizon behavior of the mode
functions $w_{L,R}$ and $w_\parallel$ in Eqs.~(\ref{eq24}) and (\ref{eq11}) we
obtain
\[
\langle\mbox{\boldmath$\xi$}_{W_{L,R}}(t)\,\mbox{\boldmath$\xi$}_{W_{L,R}}
(t')\rangle\simeq\frac{H^3}{4\pi^2}\,\delta(t-t')
\]
for the transverse noise source and
\[
\langle\mbox{\boldmath$\xi$}_{W_\parallel}(t)\,\mbox{\boldmath$\xi$}_{W_\parallel}
(t')\rangle\simeq\frac{H^3}{4\pi^2}\left(\frac{3H}{M}\right)^2\,
\delta(t-t')
\]
for the longitudinal one.

\subsection{Fokker-Planck equation}\label{sec:FPeq}
In this section we obtain and solve the Fokker-Planck equations that follow
from Eq.~(\ref{eq32}). Given such a stochastic differential equation, it is a
standard procedure to derive the corresponding Fokker-Planck equation
\cite{Gardiner}. To do so, we first introduce an  arbitrary basis of
orthonormal vectors $\mbox{\boldmath $u$}_i$ ($i=1,2,3$) in  position-space. In
such basis, the components
$W_i^\lambda\equiv\mbox{\boldmath$W$}_\lambda\cdot \mbox{\boldmath$u$}_i$ of the
$\lambda$-polarised vector are
\begin{equation}
W_i^\lambda(t)=\int\frac{d^3k}{(2\pi)^
3}\;\theta(k_s-k)\left[\mbox{\boldmath $u$}_i\cdot\mbox{\boldmath$e$}_\lambda
(\hat{\mbox{\boldmath $k$}})\,\hat a_\lambda(\mbox{\boldmath $k$})w_\lambda e^{i\mbox{\scriptsize\boldmath$k$}\cdot \mbox{\scriptsize
\boldmath$x$}}+\mbox{\boldmath $u$}_i\cdot\mbox{\boldmath$e$}_\lambda^*(\hat{\mbox{\boldmath $k$}})\,\hat a_\lambda^\dagger(\mbox{\boldmath $k$})
w_\lambda e^{i\mbox{\scriptsize\boldmath$k$}\cdot \mbox{\scriptsize
\boldmath$x$}}\right]\,.
\end{equation}
It can be shown that the operators $W_{1,2,3}^\lambda$ are formally the same, and
therefore we can write the vector \mbox{\boldmath$W$}$_\lambda$ in terms of a
scalar-like operator $W_\lambda$ defined as follows
\[\label{eq36}
\mbox{\boldmath$W$}_\lambda\equiv W_{\lambda}\,(1,1,1)\,.
\]
The expectation value of $W_\lambda$ thus determines the modulus of the
$\lambda$-polarised vector $\mbox{\boldmath$W$}_\lambda$. Of course, the same
applies to the stochastic source \mbox{\boldmath$\xi$}$_{W_\lambda}$, which can
be expressed as
\[\label{eq34}
\mbox{\boldmath$\xi$}_{W_\lambda}\equiv\xi_{W_\lambda}\,(1,1,1)
\]
after introducing the scalar-like noise $\xi_{W_\lambda}$. Using the above, the
vector Eq.~(\ref{eq32}) can be rewritten as
\[\label{eq37}
\dot W_{\lambda}+{\cal F}_\lambda(t)W_{\lambda}\simeq\xi_{W_\lambda}\,.
\]
In Appendix~\ref{appB} we compute the mean-square field using the solution to
Eq.~(\ref{eq37}) and compare the result with the obtained by solving the
Fokker-Planck equation in this section.

Similarly to Eq.~(\ref{eq84}), to determine the magnitude of the
self-correlation for the scalar-like sources $\xi_{W_{\lambda}}$ we introduce the
diffusion coefficients $D_\lambda(t)$ as follows
\[\label{eq35}
\langle\xi_{W_\alpha}(t)\,\xi_{W_\beta}(t')\rangle=D_\alpha(t)\,
\delta_{\alpha\beta}\delta(t-t')\,.
\]
Taking into account (\ref{eq34}) and comparing Eqs.~(\ref{eq84}) and
(\ref{eq35}) we find
\[\label{eq72}
D_\lambda(t)=\frac13\,{\cal D}_\lambda(t)\,,
\]
which results in the following transverse and longitudinal coefficients
\[\label{eq31}
D_{L,R}=\frac13\,\frac{H^3}{4\pi^2}\quad,\quad D_\parallel(t)=\frac13\,
\frac{H^3}{4\pi^2}\left(\frac{3H}{M}\right)^2\,.
\]

The Fokker-Planck equation that follows from Eqs.~(\ref{eq37}) and (\ref{eq35})
is \cite{Gardiner}
\begin{equation}\label{eq57}
\frac{\partial\rho_{\lambda}}{\partial t}=
\frac{\partial}{\partial W_{\lambda}}
\left({\cal F}_\lambda W_{\lambda}\rho_{\lambda}\right)
+\frac12 \frac{\partial^2}{\partial W_{\lambda}^2}
\Big(D_\lambda\,\rho_{\lambda}\Big)\,,
\end{equation}
which determines the probability density $\rho_\lambda(W_\lambda,t)$ for the
expectation value of the scalar-like operator $W_\lambda$, and hence the modulus
of $\mbox{\boldmath$W$}_\lambda$. The solution to Eq.~(\ref{eq57}) can be
readily obtained by Fourier transforming $\rho_\lambda$. Using
\[\label{eq46}
\rho_\lambda(W_\lambda,t)=\int_{-\infty}^{\infty} e^{isW_\lambda}
\widetilde{\rho_\lambda}(t,s)\,ds\,,
\]
the equation for $\widetilde{\rho_\lambda}$ is
\[\label{eq43}
\dot{\widetilde{\rho_\lambda}}=-{\cal F}_\lambda(t)\,s\,
\partial_s\widetilde{\rho_\lambda}-\frac{s^2}{2}\,D_\lambda(t)
\widetilde{\rho_\lambda}\,.
\]
To solve this equation we consider that the expectation value of $W_\lambda$
begins sharply peaked around the value \mbox{$W_\lambda=W_\lambda(0)$}, which
translates into $\rho_\lambda(W_\lambda,0)=\delta(W_\lambda-W_\lambda(0))$. Imposing
such condition, the solution to Eq.~(\ref{eq43}) is
\[\label{eq44}
\widetilde{\rho_\lambda}(t)=\frac{1}{2\pi}
\exp\left[-i\mu_\lambda(t)s-\frac{\sigma_\lambda^2(t)}{2}s^2\right]
\]
where
\[\label{eq41}
\mu_\lambda(t)\equiv W_\lambda(0)\exp\left[-\int_0^t
{\cal F}_\lambda(\tau)d\tau\right]
\]
and
\[\label{eq49}
\sigma_\lambda^2(t)\equiv\int_0^t\exp\left[-2\int_{\bar\tau}^t
{\cal F}_\lambda(\tau)d\tau\right]D_\lambda(\bar\tau)d\bar\tau\,.
\]
Integrating now Eq.~(\ref{eq46}) we find
\[\label{eq56}
\rho_\lambda(W_\lambda,t)=\frac{1}{\sqrt{2\pi}\,\sigma_\lambda(t)}\,
\exp\left[-\frac{(W_\lambda-\mu_\lambda(t))^2}{2\sigma_\lambda^2(t)}\right]\,,
\]
i.e. a Gaussian distribution with mean $\mu_\lambda(t)$ and
variance $\sigma_\lambda^2(t)$. Therefore,
\[\label{eq47}
\langle W_\lambda\rangle=\mu_\lambda(t)\quad\textrm{and}\quad
\langle W_\lambda^2\rangle=\mu_\lambda^2(t)+\sigma_\lambda^2(t)\,.
\]

\subsubsection{Transverse vector \mbox{\boldmath$W$}$_\perp$}
Using ${\cal F}_{L,R}=\frac{M^2}{9H}$ and
$D_{L,R}=\frac13\,\frac{H^3}{4\pi^2}$ for the transverse polarisations we find
\[\label{eq21}
\mu_{L,R}(t)=\exp\left[-\frac{\left(1-a^{-6}\right) M^2}{54 H^2}\right]
W_{L,R}(0)\simeq \exp\left(-\frac{M^2}{54 H^2}\right)W_{L,R}(0)
\]
and
\[
\sigma_{L,R}^2(t)=\frac{H^2\exp\left(-\frac{M^2}{27 H^2}\right)
\left[Ei\left(\frac{M^2}{27 H^2}\right)-Ei\left(\frac{M_0^2}{27 H^2}\right)
\right]}{72 \pi ^2}\,,
\]
where $Ei(x)=-\int_{-x}^\infty t^{-1}e^{-t}dt$ is the exponential integral
\cite{Abramowitz}. When $M^2\ll H^2$, using the expansion
\mbox{$Ei(x>0)=\gamma+\ln\,x+\sum_{n=1}^\infty\frac{x^n}{n\,n!}$} we find
\[\label{eq22}
\mu_{L,R}\simeq W_{L,R}(0)\quad,\quad\sigma_{L,R}^2(t)\simeq\frac{H^3t}{12\pi^2}\,,
\]
whereas for $M^2\gg H^2$, using the asymptotic expansion
\mbox{$Ei(x)\simeq\frac{e^x}{x}\sum_{n=0}^\infty\frac{n!}{x^n}$} for $x\gg1$
\cite{Abramowitz} we obtain
\[\label{eq86}
\mu_{L,R}\simeq 0\quad,\quad\sigma_{L,R}^2(t)\simeq\frac{H^2}{72\pi^2}
\left(\frac{27H^2}{M^2}\right)\ll H^2\,,
\]
Using Eq.~(\ref{eq36}) and summing over polarisations we can translate the
above results in terms of the transverse vector
$\mbox{\boldmath$W$}_\perp\equiv\mbox{\boldmath$W$}_L+\mbox{\boldmath$W$}_R$.
Considering the case $M^2\ll H^2$ only we have
\[\label{eq26}
\langle\mbox{\boldmath$W$}_\perp\rangle\simeq\mbox{\boldmath$W$}_\perp(0)\quad,
\quad\langle\mbox{\boldmath$W$}_\perp^2\rangle\simeq\mbox{\boldmath$W$}_\perp^2(0)
+2\left(\frac{H^3t}{4\pi^2}\right)\,.
\]

Although the variance computed in Eq.~(\ref{eq22}) grows linearly with time for
$M^2\ll H^2$, our result does not feature an asymptotic value corresponding to
an equilibrium state, known to exist in the case of light scalar fields
\cite{BD}. Since the transverse modes of the massive vector field behave like a
light scalar field, an analogous equilibrium value might be expected. The
reason for its non-appearance in Eqs.~(\ref{eq22}) is clear. Firstly, the
effective mass of the physical vector field grows as $M\propto a^3$. And
secondly, inflation can proceed even if $M\gg H$ since the vector field does
not play the role of inflaton (and in any case, its energy density is kept
constant). If inflation continues after $M^2\sim H^2$, the rapid scaling of $M$
makes $M^2\gg H^2$ in less than one $e$-folding. Consequently, at sufficiently
long times the field fluctuations do not approach any asymptotic value, but
undergo exponential suppression as shown in Eq.~(\ref{eq86}).

Although no equilibrium fluctuation appears in Eq.~(\ref{eq22}), it is
instructive to compare the mean-square field $\langle W_{L,R}^2\rangle$ with the
\textit{instantaneous} equilibrium value. By such instantaneous equilibrium we
refer to the asymptotic value which the mean-square would feature for a certain
value of $M$, namely
\mbox{$\langle W_{L,R}^2\rangle_{\rm eq}=\frac{H^4}{8\pi^2M^2}\propto a^{-6}$}
(see Eq.~(\ref{eq62})). At a given time $t$ before the end of inflation we have
\[\label{eq65}
\frac{\langle W_{L,R}^2\rangle_{\rm eq}}{\langle W_{L,R}^2\rangle}\sim
\left(\frac{H}{M_e}\right)^2\frac{e^{6N_e}}{N}\,,
\]
where $M_{\rm e}$ is the effective mass at the end of inflation, $N$ is the
number of $e$-foldings elapsed since the beginning of inflation and $N_e$ the
number of $e$-foldings remaining until the end of inflation. If $M_e\sim H$, at
the end of inflation $N$ corresponds to the number of the total inflationary
$e$-foldings $N=N_{\rm tot}$ and the mean-square becomes
\mbox{$\langle W_{L,R}^2(t_e)\rangle\sim N_{\rm tot}\langle
W_{L,R}^2(t_e)\rangle_{\rm eq}$}, and hence much larger than the equilibrium value
corresponding to the field's effective mass at the end of inflation, $M_e$.
Although this result may seem surprising, it clearly follows because
$\langle W_{L,R}^2\rangle\propto N$ [c.f. Eqs.~(\ref{eq47}) and (\ref{eq22})]
while the scaling of $M$ makes the instantaneous equilibrium value decrease as
$a^{-6}$. Provided $M_e$ is sufficiently close to $H$, only a moderate amount of
inflation is needed for $\langle W_{L,R}^2\rangle$ to be above its instantaneous
equilibrium value by the end of inflation. On the contrary, if $M_e\ll H$, the
mean-square $\langle W_{L,R}^2\rangle$ remains well below its final equilibrium
amplitude unless an exponentially large number of $e$-foldings is considered.
This may be the case if eternal inflation is considered \cite{Linde:1986fc}.
Finally, if $M^2\gg H^2$ during inflation the condensate becomes exponentially
suppressed very quickly, as indicated in Eq.~(\ref{eq86}).

\subsubsection{Longitudinal vector \mbox{\boldmath$W$}$_\parallel$}
Using now ${\cal F}_\parallel=3H$ and \mbox{$D_{L,R}=\frac13\,\frac{H^3}{4\pi^2}\left(\frac{3H}{M}\right)^2$} for the longitudinal polarisation we find
\[
\mu_\parallel(t)=\frac{M_0}{M}\,W_\parallel(0)
\]
and
\[\label{eq87}
\sigma_\parallel^2(t)=\frac13\left(\frac{3H}{M}\right)^2\frac{H^3t}{4\pi^2}\,.
\]
Using Eqs.~(\ref{eq36}) and  (\ref{eq47}) to translate the above results in
terms of the longitudinal vector \mbox{\boldmath$W$}$_\parallel$ we find
\[\label{eq18}
\langle\mbox{\boldmath$W$}_\parallel\rangle\simeq\frac{M_0}{M}\,
\mbox{\boldmath$W$}_\parallel(0)
\]
and
\[
\langle\mbox{\boldmath$W$}_\parallel^2\rangle\simeq\left(\frac{M_0}{M}
\right)^2\mbox{\boldmath$W$}_\parallel^2(0)
+\left(\frac{3H}{M}\right)^2\frac{H^3t}{4\pi^2}\,.
\]

Similarly to the transverse field, owing to the scaling of $M$ the variance in
Eq.~(\ref{eq87}) does not exhibit an asymptotic equilibrium value. Comparing
the mean-square $\langle W_\parallel^2\rangle$ (obtained from Eq.~(\ref{eq47}))
with its instantaneous equilibrium amplitude
\mbox{$\langle W_\parallel^2\rangle_{\rm eq}=
\left(\frac{3H}{M}\right)^2\frac{H^4}{8\pi^2M^2}$} (see Eq.~(\ref{eq63})) we
obtain
\[\label{eq66}
\frac{\langle W_\parallel^2\rangle_{\rm eq}}{\langle W_\parallel^2\rangle}\sim
\left(\frac{H}{M_e}\right)^2\frac{e^{6N_e}}{N}\,,
\]
after neglecting the pre-inflationary fluctuation in the longitudinal vector.
Since this equation coincides with (\ref{eq65}) the conclusions that apply
for the transverse vector (see below Eq.~(\ref{eq65})) are also valid for the
longitudinal vector.

\section{Other cases of interest}\label{sec_cases}
\subsection{Scale invariant spectrum with $f\propto a^2$ and
$m\propto a$}\label{sec_a}
A nearly scale-invariant spectrum of superhorizon perturbations can also be
achieved provided the kinetic function $f$ and the mass $m$ vary as
\cite{Dimopoulos:2009am}
\[
f\propto a^2\quad{\rm and}\quad m\propto a\,,
\]
which corresponds to $\alpha=2$ and $\beta=1$. In this case, the effective mass
$M$ remains constant.\footnote{This case cannot correspond to a gauge field,
since, were it the case, the kinetic function would be inversely proportional to
the gauge coupling \mbox{$f\propto e^{-2}$}, which would render the theory
strongly coupled during inflation, as \mbox{$f=1$} at the end. However, note
that a massive Abelian vector boson does not need necessarily to be a gauge
field as it is renormalisable cite{tikto}.}

The equation that follows from Eq.~(\ref{eq19}) for the transverse mode is
\[\label{eq88}
\ddot w_{L,R}+3H\dot w_{L,R}+
\left(\frac{k^2}{a^2}+M^2\right)w_{L,R}=0\,.
\]
Imposing that $w_{L,R}$ matches the Bunch-Davies (BD) vacuum solution in the
subhorizon regime $k/aH\to\infty$ we obtain
\[
w_{L,R}=a^{-3/2}\sqrt{\frac{\pi}{4H}}\,e^{i\pi(\nu+1/2)/2}H^{(1)}_\nu(k/aH)\,,
\]
where $\nu^2=9/4+M^2/H^2$. On the other hand, the longitudinal mode function
satisfies
\[
\ddot w_\parallel+\left(3+\frac2{1+r^2}\right)H\dot w_\parallel+
\left(\frac{k^2}{a^2}+M^2\right)w_\parallel=0\,,
\]
which coincides with Eq.~(\ref{eq88}) when $r\gg1$, attained on superhorizon
scales. However, on superhorizon scales we have
\mbox{$w_\parallel\simeq-\frac{H}{\sqrt{2}k^{3/2}}\frac{3H}{M}$} for $M\ll 3H$,
thus remaining approximately constant. Nevertheless, the amplitude of
$w_\parallel$ is larger than the transverse function $w_{L,R}$ by a factor of
$\frac{3H}{M}\gg1$, hence $\mbox{\boldmath$W$}_c$ is approximately
longitudinal. Also on superhorizon scales we find
\[\label{eq59}
\dot w_\lambda\simeq-\frac{M^2}{3H}\,w_\lambda\quad,
\quad \ddot w_\lambda\simeq\left(\frac{M}{3H}\right)^2M^2 w_\lambda\,,
\]
where $\lambda$ labels now any of the three polarisations.

Using the above and proceeding similarly to the previous case we find
\mbox{${\cal F}_\lambda=\frac{M^2}{3H}$} in Eq.~(\ref{eq37}), whereas the
diffusion coefficients are the same as in Eq.~(\ref{eq31}). Using now
Eqs.~(\ref{eq41}) and (\ref{eq49}) we obtain the mean field
\[
\mu_\lambda(t)=W_\lambda(0)\,e^{-\frac{M^2t}{3H}}
\]
for each polarisation, and the variances\footnote{Note that since the variances
in Eqs.~(\ref{eq62}) and (\ref{eq63}) refer to one of the three possible
polarisations of the massive vector field, a factor of three is missing with
respect to the scalar field case, for which
$\langle\phi^2\rangle=\frac{3H^4}{8\pi^2 m_\phi^2}$ in the equilibrium state
\cite{BD}.}
\[\label{eq62}
\sigma_{L,R}^2=\frac{H^4}{8\pi^2M^2}\left(1-e^{-\frac{2M^2t}{3H}}\right)\,
\]
and
\[\label{eq63}
\sigma_\parallel^2=\frac{H^4}{8\pi^2M^2}\left(\frac{3H}{M}\right)^2
\left(1-e^{-\frac{2M^2t}{3H}}\right)\,.
\]
At sufficiently early times $M^2t\ll H$, the computed variances grow linearly
with time, approaching their equilibrium amplitude at late times
$M^2t\gg H$. As discussed in Sec.~{\ref{sec:FPeq}}, the equilibrium amplitude
becomes apparent now thanks to the constancy of $M$.

\subsection{Massless vector field}\label{sec_mless}
We consider a massless vector field with a time-dependent Maxwell term
\[
{\cal L}=-\frac14 f F_{\mu\nu}F^{\mu\nu}\,,
\]
with $f\propto a^\alpha$. Systems similar to this have been studied in
\cite{Martin:2007ue} and have been extensively considered for the formation of
a primordial magnetic field \cite{ratra}, or the creation of a vector field
condensate in order to render inflation mildly anisotropic \cite{anisinf}.%
\footnote{The theory is gauge invariant so it is questionable whether
a condensate is physical, as one can always add to the vector field an arbitrary
vector constant \mbox{$\mbox{\boldmath $W$}\rightarrow\mbox{\boldmath $W$}+
\mbox{\boldmath $C$}$}.}

The equation for the potential vector field is
\[
\ddot{\!\!\mbox{\boldmath$A$}}+\left(H+\frac{\dot f}{f}\right)
\dot{\!\mbox{\boldmath$A$}}-a^{-2}\nabla^2\mbox{\boldmath$A$}
=0\,,
\]
which follows from Eq.~(\ref{eq54}) after taking $m=0$ and $A_t=0$. Since a
massless vector field has two physical degrees of freedom only, the sum in
Eq.~(\ref{eq68}) runs over transverse polarisations, i.e. $\lambda=L,R$. Also,
the commutation relations satisfied by the transverse modes are as in
Eq.~(\ref{eq29}).

Using $M=0$ and $A_t=0$ in Eq.~(\ref{eq19}) we obtain the equation of motion for the transverse modes of the physical vector field
\[
\ddot w_{L,R}+3H\dot w_{L,R}+\left(m_{\rm eff}^2+\frac{k^2}{a^2}\right)w_{L,R}=0\,,
\]
where $m_{\rm eff}^2\equiv-\frac14(\alpha+4)(\alpha-2)H^2$. Demanding that
$w_{L,R}$ matches the BD vacuum solution in the subhorizon limit, the solutions
to the above equation are \cite{Dimopoulos:2006ms}
\[\label{eq71}
w_{L,R}=a^{-3/2}\sqrt{\frac{\pi}{4H}}\,e^{i\pi(\nu+1/2)/2}H^{(1)}_\nu(k/aH)\,,
\]
where now $\nu^2=9/4-m_{\rm eff}^2/H^2=(1+\alpha)^2/4>0$. For the particular
values $\alpha=-4,2$, both corresponding to $\nu=3/2$, a flat perturbation
spectrum follows. Allowing $\alpha$ to take on any other value we find
\[\label{eq69}
\dot w_{L,R}\simeq\left(\nu-\frac32\right)H\,w_{L,R}\quad,
\quad\ddot w_{L,R}\simeq\left(\nu-\frac32\right)^2H^2\,w_{L,R}
\]
in the superhorizon regime $k/aH\to0$.

Performing the coarse-graining of the physical vector field, the equation of
motion for $\mbox{\boldmath$W$}_{L,R}$ is given by Eq.~(\ref{eq1}) with $M^2$
replaced by $m_{\rm eff}^2$. Using the expression for $\dot w_{L,R}$ in
Eq.~(\ref{eq69}) to compute $\mbox{\boldmath$\Pi$}_{L,R}$ (see
Eq.~(\ref{eq70})) and comparing with Eq.~(\ref{eq32}) we obtain
\mbox{${\cal F}_{L,R}=\left(\frac32-\nu\right)H$}. Using Eqs.~(\ref{eq85}) and
(\ref{eq71}) we find the diffusion coefficient
\[
{\cal D}_{L,R}=\left(\frac{H^3}{4\pi^2}\right)\frac{4^{\nu-1/2}\Gamma^2(\nu)
\,\epsilon^{3-2\nu}}{\pi}
\,.
\]

We note that the above grows unbounded as $\nu$ increases. This is because for
large $\nu^2\gg 1$ the physical field becomes tachyonic in the subhorizon
regime. Therefore, by the time of horizon exit the amplitude $w_\lambda$ has
grown exponentially. Using Eqs.~(\ref{eq72}), (\ref{eq41}) and (\ref{eq49}) we
obtain the mean field
\[\label{eq74}
\mu_\lambda(t)=e^{(\nu-3/2)Ht}W_\lambda(0)
\]
and the variance
\[\label{eq75}
\sigma_\lambda^2(t,k)=\frac{2^{-3+2\nu}
\left[1-e^{-H t(3-2\nu)}\right]H^2\Gamma(\nu)^2
\epsilon ^{3-2\nu}}{\pi ^3 (3-2 \nu )}\,.
\]
The case of an exactly massless field for which
$\sigma_\lambda^2=\frac{H^3t}{4\pi^2}$ is trivially recovered in the limit
$\nu\to3/2$.

\subsection{Non-minimally coupled vector field}
We consider now a vector field non-minimally coupled to gravity. This theory
has been studied, for example, in Refs.~\cite{nonmin,Dimopoulos:2008yv}. In
Ref.~\cite{peloso} the theory has been criticised for giving rise to ghosts,
corresponding to the longitudinal perturbations, when subhorizon. However, the
existence of ghosts and their danger to the stability of the theory is still
under debate, see for example Ref.~\cite{David+Mindaugas}.

Consider the Lagrangian density
\[
{\cal L}=-\frac14 F_{\mu\nu}F^{\mu\nu}+\frac12m^2A_\mu A^\mu+\frac12\,
\gamma R A_\mu A^\mu\,,
\]
where $R$ is the Ricci scalar and $\gamma$ is a constant. During de Sitter
inflation $R\simeq-12H^2$, and the effective mass for the vector field is
\[
m_{\rm eff}^2=m^2+\gamma R\simeq m^2-12\gamma H^2\,.
\]

After taking $f=1$ and substituting $m^2\to m_{\rm eff}^2$, the former Lagrangian
density can be considered a special case of the more general Lagrangian density
in Sec.~\ref{sec_case}, which then simplifies the study of the condensate
formation. The equation for the vector field \mbox{\boldmath$A$} is
\[
\ddot{\!\!\mbox{\boldmath$A$}}+H\dot{\!\mbox{\boldmath$A$}}+m^2_{\rm eff}
\mbox{\boldmath$A$}-a^{-2}\nabla^2\mbox{\boldmath$A$}
=-2H\nabla A_t
\]
and the evolution for the perturbation $\delta\mbox{\boldmath$W$}$ is obtained
by replacing \mbox{$M^2\to m_{\rm eff}^2$} and taking $\alpha=0$ in
Eq.~(\ref{eq19}).

\subsubsection{Transverse modes}
For the transverse mode functions $w_{L,R}$ we have
\[\label{eq76}
\ddot w_{L,R}+3H\dot w_{L,R}+\left(2H^2+m^2_{\rm eff}+\frac{k^2}{a^2}\right)w_{L,R}
=0\,,
\]
whose solution, while matching the vacuum in the subhorizon limit, is given by
Eq.~(\ref{eq71}) with $\nu^2\equiv1/4-m_{\rm eff}^2H^2$.

With $\gamma\approx0$, the perturbation spectrum is ${\cal P}_{L,R}\propto k^2$
when $m\ll H$, thus reproducing the vacuum value. On the other hand, if
$m\gg H$, the vector is a heavy field and the buildup of fluctuations becomes
suppressed. Only when $m^2\approx-2H^2$ can the vector field be substantially
produced during inflation \cite{Dimopoulos:2006ms}. In such case, the evolution
of $w_{L,R}$ is determined by Eq.~(\ref{eq71}) with $\nu\approx3/2$ (which
corresponds to either $\alpha\approx-4$ or $\alpha\approx2$). When
$\gamma\neq0$, the vector field obtains a flat perturbation spectrum
($\nu\approx3/2$) provided $\gamma$ is tuned according to
\[
\gamma\approx \frac16\left(1+\frac{m^2}{2H^2}\right)\,.
\]
In both cases, the mode functions $w_{L,R}$ satisfy Eq.~(\ref{eq69}) in the
superhorizon regime. Consequently, the mean and variance of the transverse
vector condensate are given by Eqs.~(\ref{eq74}) and (\ref{eq75}) using
$\nu^2=1/4-m_{\rm eff}^2H^2$.

\subsubsection{Longitudinal modes}
The evolution equation for the longitudinal modes is
\[\label{eq78}
\ddot w_\parallel+\left(3+\frac{2k^2}{k^2+a^2m_{\rm eff}^2}\right)H\dot w_\parallel+
\left(2H^2+m^2_{\rm eff}+\frac{2H^2k^2}{k^2+a^2m_{\rm eff}^2}+\frac{k^2}{a^2}\right)
w_\parallel=0\,.
\]
If $k^2\ll a^2m_{\rm eff}^2$ during inflation, this equation becomes identical to
Eq.~(\ref{eq76}). Consequently, the results that apply for the transverse field
are valid for the longitudinal component too. In the opposite regime
($k^2\gg a^2m_{\rm eff}^2$) the vector field indeed obtains a flat perturbation
spectrum. When \mbox{$0<m_{\rm eff}^2\ll H^2$} satisfying the condition
$k^2\gg a^2m_{\rm eff}^2$ approximates \mbox{$m_{\rm eff}^2\approx 0$}, when the
longitudinal component decouples from the theory and is unphysical.
However, if $m_{\rm eff}^2\approx -2H^2$ the longitudinal component can be
produced while attaining a flat perturbation spectrum. Writing
$m_{\rm eff}^2=-2H^2$, the solution to Eq.~(\ref{eq78}) is
\cite{Dimopoulos:2008yv,Dimopoulos:2011ws}
\[\label{eq80}
w_\parallel=\frac{e^{ik/aH} H}{2 k^{3/2}}
\left(-2+2i\frac{k}{aH}+\frac{k^2}{a^2H^2}\right)\,.
\]
Deriving and taking the limit $k/aH\to0$ we find\footnote{Using
Eq.~(\ref{eq80}) it can be checked that $\ddot w_\parallel$ exactly satisfies
$\ddot w_\parallel+\left(3+\frac{ik}{aH}\right)H\dot w_\parallel=0$, which can
be used to rewrite Eq.~(\ref{eq78}) as a first order equation
\[
\left(\frac{2k^2}{k^2-2a^2H^2}-\frac{ik}{aH}\right)H\dot w_\parallel+
\left(\frac{2H^2k^2}{k^2-2a^2H^2}+\frac{k^2}{a^2}\right)w_\parallel=0\,.
\]}
\[
\dot w_\parallel\simeq\frac{i}{2}\left(\frac{k}{aH}\right)^3Hw_\parallel\quad,
\quad\ddot w_\parallel\simeq-3H\dot w_\parallel\,.
\]
Substituting now $\dot w_\parallel$ in Eq.~(\ref{eq70}) gives rise to gradient
terms in Eq.~(\ref{eq82}). After neglecting these, Eq.~(\ref{eq32}) becomes
\[
\mbox{\boldmath$\dot W$}_\parallel\simeq\mbox{\boldmath $\xi$}_{W_\parallel}(t)\,,
\]
hence ${\cal F}_\parallel=0$. Using that $w_\parallel\simeq-\frac{H}{k^{3/2}}$ on
superhorizon scales (implying ${\cal P}_\parallel=2{\cal P}_{L,R}$
\cite{Dimopoulos:2008yv,Dimopoulos:2011ws}) we find the diffusion coefficient
$D_\parallel=2D_{L,R}=\frac{1}{3}\frac{H^3}{2\pi^2}$. Finally, using
Eqs.~(\ref{eq72}), (\ref{eq41}) and (\ref{eq49}) we obtain the mean field and
variance
\[
\mu_\parallel(t)= W_\lambda(0)\quad,\quad\sigma_\parallel^2(t)=\frac13\frac{H^3t}{2\pi^2}\,.
\]

\subsection{Parity violating vector field}
Recently, a parity violating, massive vector field has been considered in
the context of the vector curvaton mechanism \cite{Dimopoulos:2012av} in the
effort to generate parity violating signatures on the microwave sky (see also
Refs.~\cite{sorbo,gaugeflation}).

The Lagrangian density considered is
\[
{\cal L}=-\frac14fF_{\mu\nu}F^{\mu\nu}
-\frac14hF_{\mu\nu}\tilde F^{\mu\nu}+\frac12m^2A_\mu A^\mu\,.
\]
Since the axial term does not affect the longitudinal component of the
perturbation \cite{Dimopoulos:2012av}, taking $f\propto a^{-1\pm3}$ and
$m\propto a$ we obtain a scale invariant perturbation spectrum for the
longitudinal component with amplitude
\[
{\cal P}_\parallel=\left(\frac{3H}{M}\right)^2\left(\frac{H}{2\pi}\right)^2\,.
\]

The equation of motion for the transverse polarisations is
\[\label{eq50}
\ddot w_\pm+3H\dot w_\pm+\left(\frac{k^2}{a^2}+M^2\pm Q^2\right)w_\pm=0
\]
where $Q^2\equiv\frac{k}{a}\frac{|\dot h|}{f}$. If $\dot h$ is positive during inflation the subscripts $+\equiv R$ and $-\equiv L$, whereas $+\equiv L$ and $-\equiv R$ if $\dot h$ is negative.  We focus on the case when $Q^2$ is the dominant term in the above brackets on superhorizon scales. The case when $M^2$ is the dominant term is studied in \cite{Dimopoulos:2009am}. We further assume that $M^2$ never dominates in the subhorizon regime. Upon
parametrising the time-dependence of $Q^2$ as $Q\propto a^c$, the equation
\[
\ddot w_\pm+3H\dot w_\pm\pm Q^2w_\pm=0
\]
can be solved exactly. The arbitrary constants in the general solution are
chosen so that $w_\pm$ and $\dot w_\pm$ match the BD vacuum solution in the
subhorizon limit $k/aH\to\infty$.

In order to describe the evolution of an individual mode, four cases are
identified according to the magnitude of $Q/H$ during inflation. In what
follows, we illustrate the buildup of the condensate when $Q_e\ll H$. Of
course, when $Q_e\gg H$ the amplitude of the vector fluctuations are
suppressed. Regarding the evolution of $w_\pm$ subject to the condition
$Q_e\ll H$, two cases can be considered:

\begin{itemize}
  \item {\bf Case I:} $Q\ll H$ at all times.
Defining the scale factor $a_X$ by the condition $k/a_X=Q$,  and $a_*$ by
$k/a_*=H$ (horizon crossing), the condition $Q\ll H$ can be rewritten as $a_*\ll a_X$.
This implies that the mode functions $w_\pm$ behave as a light field at all times during inflation. To zero order in $Q/H$, the growth of fluctuations proceeds as if the parity violating term were absent. Parity violating effects appear as higher order corrections in $Q/H$, which can be neglected to estimate the magnitude of the condensate at the end of inflation. To zero order in $Q/H$, the power spectrum for these modes is
      \[\label{eq60}
      {\cal P}^{\rm (I)}_{w_\pm}(k)=\left(\frac{H}{2\pi}\right)^2\,.
      \]

  \item {\bf Case II:} $Q\gg H$ during an earliest stage of inflation,
but $Q_e\ll H$. In this case, the mode function $w_\pm$ behaves as follows: For $a<a_X$, the mode functions $w_\pm$ approach the BD vacuum solution, thus behaving as modes of an effectively massless field. For $a_X<a<a_*$, $w_\pm$ behave as modes of a heavy field. Consequently, the amplitude of the vacuum fluctuations at horizon crossing is suppressed. For $a_*<a<a_H$, where $a_H$ is defined by $Q=H$, the modes continue behaving as those of a heavy field, thus oscillating and reducing the amplitude of their vacuum fluctuation. During the final stage of inflation: $a_H<a<a_e$, the mode ceases to oscillate and obtains an expectation value. If the previous phase of oscillations is long-lasting, and depending on the value of $c$, the amplitude of the mode can become very suppressed by the end of inflation. To order zero in $Q/H$, the
power spectrum for these modes is scale independent when $c=-1/2$, which can be
naturally realised when string axions are considered \cite{Dimopoulos:2012av}:
      \[\label{eq61}
      {\cal P}^{\rm (II)}_{w_+}=\frac{4}{\pi}\left(\frac{Hf}{\dot h}\right)^3
\left(\frac{H}{2\pi}\right)^2\,\,,\,\,{\cal P}^{\rm (II)}_{w_-}=
\frac12\,{\cal P}^{\rm (II)}_{w_+}\,\exp\left(\frac{4\dot h}{Hf}\right)\,.
      \]
\end{itemize}

Although cases I and II describe the evolution of a single mode, the condensate
formed during inflation contains a collection of modes which can span many
orders of magnitude in momentum space. Consequently, in the most general case
the condensate encompasses modes which have undergone different evolution, and
therefore their amplitudes can be much different. For example, if $Q\ll H$
at all times during inflation, the evolution of the modes in the condensate is dictated by case~I only. Nevertheless, if $Q\gg H$ initially, the condensate at the end of inflation is
made up of modes with evolution dictated by case~II (modes exiting the horizon before
$Q=H$) and by case~I (modes exiting the horizon after $Q=H$). This is in contrast to the cases previously studied, for which all the modes in the condensate undergo the same evolution.

To compute the mean square field we simply add up the square amplitude of the
modes that are superhorizon at the end of inflation and disregard the contribution from modes that are superhorizon at the beginning.  Bearing in mind the foregoing discussion and using Eqs.~(\ref{eq60}) and (\ref{eq61}) we find
\[
\langle\mbox{\boldmath$W$}_+^2\rangle\simeq\frac13
\left(\int_{H}^{He^{N_{\rm II}}}{\cal P}^{\rm (II)}_{w_+}(k)\,\frac{dk}{k}+
\int_{He^{N_{\rm II}}}^{He^{N_{\rm I}}}{\cal P}^{\rm (I)}_{w_+}(k)\,\frac{dk}{k}\right)=
\frac{H^2}{12\pi^2}\left[N_{\rm I}+\frac{4N_{\rm II}}{\pi}
\left(\frac{H f_e}{\dot h_e}\right)^3\right]\,,
\]
where $N_{\rm II}$ is the number of $e$-foldings from the beginning of inflation
until $Q=H$ and $N_{\rm I}$ is the remaining number of $e$-foldings until the
end of inflation. To estimate the length of inflation while $Q>H$ we take into
account that the first mode that crosses outside the horizon during inflation
is $k_0/a_0\sim H$. Therefore, at the beginning of inflation we can estimate
$Q_0\simeq (H|\dot h_0|/f_0)^{1/2}$. Using now that $Q\propto a^c$, the number of
$e$-foldings until $Q=H$ is \mbox{$N_{\rm II}\simeq
\frac1{c}\ln\frac{H}{Q_0}\simeq\frac1{2c}\ln\frac{H f_0}{|\dot h_0|}$}. Writing
the total number of $e$-foldings as $N_{\rm tot}=N_{\rm I}+N_{\rm II}$ we have
\[
\langle\mbox{\boldmath$W$}_+^2\rangle\simeq\frac{H^2}{12\pi^2}
\left\{N_{\rm tot}+\left[\frac{4}{\pi}\left(\frac{H f_e}{\dot h_e}\right)^3-1
\right]\frac1{2c}\ln\left(\frac{Hf_0}{|\dot h_0|}\right)\right\}\,.
\]

Proceeding similarly we find the mean square for the mode $w_-$
\[
\langle\mbox{\boldmath$W$}_-^2\rangle\simeq\frac{H^2}{12\pi^2}\left\{N_{\rm tot}+\left[\frac{2}{\pi}\left(\frac{H f_e}{\dot h_e}\right)^3\exp\left(\frac{4\dot h}{Hf}\right)-1\right]\frac1{2c}\ln\left(\frac{Hf_0}{|\dot h_0|}\right)\right\}\,.
\]
The exponential amplification with respect to $\mbox{\boldmath$W$}_+^2$ is due
to the fact that, when the effective potential for $w_-$ becomes tachyonic at
$a=a_X$, the mode undergoes a fast-roll motion \cite{Linde:2001ae} until its
evolution becomes overdamped at $a=a_H$. Consequently, the amplitude of the
mode $w_-$ undergoes exponential amplification for $a_X<a<a_H$. The immediate
consequence of this fact is the subsequent exponential amplification
of the transverse vector condensate, which may well overwhelm the longitudinal
component and dominate the entire condensate.

As discussed in Ref.~\cite{Dimopoulos:2012av} (see also Ref.~\cite{fnlanis}),
parity
violating signals cannot source parity violating statistical anisotropy in the
power spectrum of the curvature perturbation, because the latter (i.e. $g$)
depends only on the even combination ${\cal P}_+$ of the transverse spectra
(c.f. Eq.~(\ref{g})). Parity violating signals appear only in higher
order correlators of the curvature perturbation such as the bispectrum,
trispectrum etc. However, the observations of the Planck satellite have not
detected any significant non-Gaussianity as yet \cite{planck+}.

From the above we see that, even though parity violation is hard to observe at
the moment in the cosmological perturbations, the parity violating axial model
can result in exponential amplification of the vector field condensate. This,
in turn, can have drastic implications on observables stemming from the
existence of such a vector condensate, such as statistical anisotropy, as we
discussed in Sec.~\ref{sec-stanis}.

\section{Classical versus quantum evolution}
We now return to the varying kinetic function and mass theory, discussed in
Sec.~\ref{sec_case}. From the results in Sec.~\ref{sec-stocha} we can obtain
the mean-square of coarse-grained vector \mbox{\boldmath$W$}$_c$. Using
Eqs.~(\ref{eq26}) and (\ref{eq18}) we have
\begin{equation}
\langle\mbox{\boldmath$W$}_c^2\rangle\simeq
\mbox{\boldmath$W$}_\perp(0)^2+\left(\frac{M_0}{M}\right)^2
\mbox{\boldmath$W$}_\parallel(0)^2
+2\,\frac{H^3t}{4\pi^2}+\left(\frac{3H}{M}
\right)^2\frac{H^3t}{4\pi^2}\,.
\end{equation}
Given that $M\ll 3H$, the coarse-grained vector is dominated by the
longitudinal modes (see e.g Eqs.~(\ref{eq24}) and (\ref{eq11})), which allow us
to disregard the contribution from the transverse modes for the most part of
inflation. Consequently, and introducing $W\equiv\mbox{\boldmath$W$}_c$ and
\mbox{$W_0\equiv\mbox{\boldmath$W$}_\parallel(0)$} for notational simplicity, for
the vector field we can write
\begin{equation}
\langle W^2\rangle=\left(\frac{M_0}{M}\right)^2W_0^2
+\left(\frac{3H}{M}\right)^2\left(\frac{H}{2\pi}\right)^2\Delta N\,,
\label{Wtotal}
\end{equation}
where \mbox{$\Delta N=H\Delta t$} denotes the elapsing $e$-foldings and
\mbox{$M\propto a^3$}. From the above equation we see that, while the
homogeneous ``zero''-mode of the vector field (square-root of first term)
scales as \mbox{$\propto a^{-3}$} during inflation (before the possible onset
of its oscillations), the region of the ``diffusion zone'' in field space,
which corresponds to the accumulated fluctuations (square-root of second
term), scales as $\propto a^{-3}\sqrt{\ln a}$, since
\mbox{$\Delta N\propto\ln a$}. This means that the diffusion zone diminishes
slightly slower than the amplitude of the ``quantum kick''
\mbox{$\delta W\sim H^2/M\propto a^{-3}$}. As a result, given enough e-folds,
the vector field condensate will assume a large value which will dominate over
subsequent ``quantum kicks''. In a sense, once the condensate is
\mbox{$W\gg\delta W$}, the ``quantum kicks'' become irrelevant to
its evolution, which follows the classical equations of motion. This is
analogous to the scalar field case. Indeed, when the scalar potential is flat,
the scalar field condensate due to the accumulated fluctuations, grows as
\mbox{$\langle\phi^2\rangle\sim H^3\Delta t\propto\Delta N$}
\cite{Linde:1982uu}, so it can, in time, become much larger than the value of
the ``quantum kick'' \mbox{$\delta\phi=H/2\pi$}.

Another consequence of the fact that the diffusion zone diminishes slower than
the zero-mode is that the initial value of the vector field condensate is, in
time, overwhelmed by the quantum diffusion contribution, and can, eventually,
be ignored. Thus, when the cosmological scales exit the horizon we can consider
only the last term in the above equation, giving
\begin{equation}
\langle W_*^2\rangle=
\left(\frac{3H}{M_*}\right)^2\left(\frac{H}{2\pi}\right)^2N_p\;,
\end{equation}
where with $N_p$ we denote the number of $e$-foldings which have passed since
the beginning of inflation until the time when the cosmological scales exit the
horizon, i.e. \mbox{$N_p=N_{\rm tot}-N_*$}, with $N_*$ being the number of
the remaining $e$-foldings of inflation when the cosmological scales leave the
horizon. 

We can now use the above value as our initial homogeneous ``zero''-mode and
follow the development of the condensate after the exit of the cosmological
scales. Employing Eq.~(\ref{Wtotal}), we find
\begin{equation}
\langle W^2\rangle=\left(\frac{M_*}{M}\right)^2W_*^2
+\left(\frac{3H}{M}\right)^2\left(\frac{H}{2\pi}\right)^2\Delta N_*=
\left(\frac{3H}{M}\right)^2\left(\frac{H}{2\pi}\right)^2(N_p+\Delta N_*)\,,
\end{equation}
where \mbox{$\Delta N_*$} denotes the elapsing e-folds {\em after} the
cosmological scales exit the horizon. Since \mbox{$\Delta N_*\leq N_*$} we
can safely assume that the amount contributed by the quantum diffusion to
$\langle W^2\rangle$ from $t_*$ can be
ignored if \mbox{$N_*<N_p$} or equivalently if \mbox{$N_{\rm tot}>2N_*$}.
This is a reasonable assumption to make, given that inflation can be
long-lasting. If this is the case then, after the cosmological scales exit the
horizon and for all intends and purposes, the value of the vector field
condensate scales as \mbox{$W\propto a^{-3}$}, while we can take
\begin{equation}
W_*\simeq
\frac{H^2}{\varepsilon M_*}\,.
\label{W*}
\end{equation}
where we have defined
\begin{equation}
\varepsilon\equiv\frac{2\pi}{3\sqrt{N_p}}\ll 1\,.
\label{eps}
\end{equation}

The value of the vector field condensate $W_*$ when the cosmological scales
exit the horizon was considered a free parameter in all studies until now,
as explained in Sec.~\ref{sec-stanis},
and results were expressed in terms of it.
However, in this paper we have managed to produce an estimate of this quantity
based on physical reasoning, which is given in Eq.~(\ref{W*}).\footnote{This is
analogous to the well-known Bunch-Davis result, where the initial condition of
a light scalar field in a quadratic potential was found to be
\mbox{$\langle\phi^2\rangle\sim H^4/m_\phi^2$}, with $m_\phi$ being the mass of
the scalar field \cite{BD}.} Using this equation,
we now investigate whether the desired observational outcomes (e.g. observable
statistical anisotropy) can be obtained with realistic values of the remaining
free parameter \mbox{$\varepsilon\sim N_p^{-1/2}$}.

The first bound we can obtain for $\varepsilon$ comes from the requirement that
the density of the vector boson should not dominate the density of
inflation.\footnote{We do not consider vector inflation here \cite{vecinf}.}
As shown in Ref.~\cite{Dimopoulos:2009am} the density of the
vector field is \mbox{$\rho_W=M^2W^2=\,$cte.} Evaluating this
at the horizon exit of the cosmological scales we have
\begin{equation}
(M_*W_*)^2=\rho_W<\rho_{\rm inf}=3H^2m_P^2
\Rightarrow\varepsilon>\frac{1}{\sqrt 3}\frac{H}{m_P}\,.
\label{epsbound}
\end{equation}
This implies that the inflationary period cannot be arbitrary large.
Indeed, the range of $N_p$ values is
\begin{equation}
N_*<N_p<\frac{4\pi^2}{3}\left(\frac{m_P}{H}\right)^2.
\label{Nprange}
\end{equation}

\section{Vector curvaton physics}

In this section we apply the above into the vector curvaton scenario following
the findings of Ref.~\cite{Dimopoulos:2009am}. We consider a massive vector
field with varying kinetic function \mbox{$f\propto a^{-4}$} and mass
\mbox{$M\propto a^3$}. The vector field is subdominant during inflation
and light when the cosmological scales exit the horizon. Afterwards, it becomes
heavy (this can occur even before the end of inflation) and undergoes coherent
oscillations, during which it behaves as pressureless and {\em isotropic}
matter \cite{Dimopoulos:2006ms}.
Hence, after inflation, its density parameter grows in time
and has a chance of contributing significantly to the curvature perturbation in
the Universe, generating for example observable statistical anisotropy. For a
review of the mechanism see Ref.~\cite{Dimopoulos:2011ws}.

\subsection{Light vector field}\label{lvf}

As before, by ``light'' we mean a vector field whose mass $M$ remains
\mbox{$M<H$} until the end of inflation. At the end of inflation we assume that
the vector field becomes canonically normalised (i.e. \mbox{$f=1$}) and $M$
assumes a constant value \mbox{$M_{\rm end}\equiv m$}. As discussed in
Ref.~\cite{Dimopoulos:2009am}, in this case the vector field
undergoes strongly anisotropic particle production so that its role can only
be to generate statistical anisotropy in the curvature perturbation $\zeta$,
while leaving the dominant contribution to the spectrum of $\zeta$ to be
accounted for by some other isotropic source, e.g. the inflaton scalar field.

In this case, the anisotropy parameter $g$, which quantifies the statistical
anisotropy in the spectrum, is related to $\zeta$ as
\cite{Dimopoulos:2009am}
\begin{equation}
\zeta\sim\frac{1}{\sqrt g}\Omega_{\rm dec}\zeta_W\;,
\label{gz}
\end{equation}
where \mbox{$\Omega_{\rm dec}\equiv(\rho_W/\rho)_{\rm dec}$} is the density
parameter of the vector field at the time of its decay and
\begin{equation}
\zeta_W \sim\left.\frac{\delta W}{W}\right|_{\rm end}
\end{equation}
is the curvature perturbation attributed to the vector field. Using that
\mbox{$\delta W=(\frac{3H}{M})(\frac{H}{2\pi})$} and that
\mbox{$M\propto W^{-1}\propto a^3$} we find
\begin{equation}
\zeta_W\sim
\varepsilon\,.
\end{equation}

In Ref.~\cite{Dimopoulos:2009am} it is shown that this
scenario generates predominantly anisotropic non-Gaussianity, which peaks in
the equilateral configuration. In this configuration, we have
\cite{Dimopoulos:2009am}
\begin{equation}
\frac65 |f_{\rm NL}^{\rm equil}|=\frac14\frac{g^2}{\Omega_{\rm dec}}\,.
\end{equation}
According to the latest Planck data \mbox{$|f_{\rm NL}^{\rm equil}|\lesssim 120$}
(at 95\% CL) \cite{planck+}.
Using this bound and Eq.~(\ref{gz}), it is easy to
find that \mbox{$g<24\sqrt{\Omega_{\rm dec}}$} and also
\mbox{$\zeta^4\gsim 10^{-3}\varepsilon^4\Omega_{\rm dec}^3$}. Combining this
with Eq.~(\ref{gz}) we obtain
\begin{equation}
g\lesssim(10^3\zeta/\varepsilon)^{2/3}\simeq 0.05 N_p^{1/3}
\label{gbound}
\end{equation}
Thus, we see that we can obtain observable statistical anisotropy in the
spectrum 
even with \mbox{$\varepsilon\sim 1$} (i.e. $N_p$ of a few),
where we saturated the non-Gaussianity bound and
used that \mbox{$\zeta=4.8\times 10^{-5}$}. From Eqs.~(\ref{epsbound}) and
(\ref{gbound}) we obtain
\begin{equation}
\frac{H}{m_P}<\varepsilon\sim\frac{1}{\sqrt{N_p}}\lesssim 10^3\zeta g^{-3/2},
\label{epsrange1}
\end{equation}
where we also considered Eq.~(\ref{eps}).
If we take statistical anisotropy to be observable
($g$ of a few per cent), then the above becomes
\begin{equation}
1\lesssim N_p<\left(\frac{m_P}{H}\right)^2,
\label{Npbound}
\end{equation}
which incorporates the entire allowed range for $N_p$ shown in
Eq.~(\ref{Nprange}). This means that observable statistical anisotropy in the
spectrum of $\zeta$ is quite possible. For example, from Eq.~(\ref{gbound}),
saturating the non-Gaussianity bound, we have
\mbox{$g_{\rm max}\sim 0.05\,N_p^{1/3}\gsim 0.05\,N_*^{1/3}$}, which is
indeed observable for \mbox{$N_*\approx 60$}.\footnote{The case when
\mbox{$f\propto a^2$} and \mbox{$H\gg M=\,$cte} also leads to scale
invariant anisotropic particle production \cite{Dimopoulos:2009am} as also
discussed in Sec.~\ref{sec_a}. In this case, we can still use Eq.~(\ref{W*}) as
the initial condition with \mbox{$M_*=M=\,$cte.} The results are identical to
the ones in Sec.~\ref{lvf}.}

\subsection{Heavy vector field}

We now consider the possibility that the final value of the mass of our vector
boson is \mbox{$m\gsim H$}. In this case, as shown in
Ref.~\cite{Dimopoulos:2009am}, particle production is rendered isotropic by the
end of inflation.\footnote{This may not be true if the variation of the kinetic
function and mass are due to a rolling scalar field, which also undergoes
particle production. Then, the cross-coupling of the vector and scalar
perturbations introduces an additional source term that may enhance
statistical anisotropy \cite{namba}. We do not consider this possibility here.}
This means that the vector field alone can generate
the observed curvature perturbation without the need for the direct
contribution of any other source such as a scalar field. The generated
curvature perturbation is \cite{Dimopoulos:2009am}
\begin{equation}
\zeta\sim\Omega_{\rm dec}\zeta_W\;.
\label{zz}
\end{equation}

The vector  field condensate can begin oscillating a few e-folds $(\lsim 4)$
before the end of inflation \cite{Dimopoulos:2009am}. In this case, we have
\begin{equation}
\zeta_W \sim\left.\frac{\delta W}{W}\right|_{\rm osc}\sim\varepsilon\,,
\label{ze}
\end{equation}
where the subscript `osc' denotes the onset of the oscillations and we
considered Eq.~(\ref{W*}) and that \mbox{$M_{\rm osc}\simeq H$}.

The generated non-Gaussianity in this case is
\cite{Dimopoulos:2009am}
\begin{equation}
f_{\rm NL}=\frac{5}{4\Omega_{\rm dec}}\,,
\end{equation}
as in the scalar curvaton case. Since observations suggest
\mbox{$|f_{\rm NL}^{\rm local}|\lesssim 8$} \cite{planck+}, we find
\mbox{$\Omega_{\rm dec}\gsim 0.1$}
Thus, because of the observed value of $\zeta$, we see that
\mbox{$\varepsilon\lesssim 10^{-4}$}.

In Ref.~\cite{Dimopoulos:2009am} it is shown that a heavy vector curvaton with
prompt reheating satisfies
\begin{equation}
\frac{H}{m_P}\gsim\sqrt{\Omega_{\rm dec}}\,\zeta_W
\left(\frac{\Gamma_W}{H}\right)^{1/4},
\end{equation}
where \mbox{$\Gamma_W$} is the vector curvaton's decay rate. Assuming that the
vector curvaton decays at least through gravitational couplings we have
\mbox{$\Gamma_W\gsim m^3/m_P^2$}, which simplifies the above to
\begin{equation}
\frac{H}{m_P}\gsim\frac{\zeta^2}{\Omega_{\rm dec}}
\sim\varepsilon^2\Omega_{\rm dec}\;,
\end{equation}
where we also used Eqs.~(\ref{zz}) and (\ref{ze}) and considered that
\mbox{$m\geq H$} for a heavy field. Using the fact that
\mbox{$\Omega_{\rm dec}\gsim 0.1$} to avoid excessive non-Gaussianity, we
obtain\footnote{This bound is further strengthened if reheating is not prompt
and \mbox{$\Gamma_W\gg m^3/m_P^2$}.}
\begin{equation}
\varepsilon\lesssim
\sqrt{\frac{H}{m_P}}\,.
\end{equation}
Thus, in view of Eq.~(\ref{epsbound}), we find
\begin{equation}
\frac{H}{m_P}<\varepsilon\sim\frac{1}{\sqrt{N_p}}
\lesssim \min\left\{10^{-4};\sqrt{\frac{H}{m_P}}\right\}.
\label{epsrange2}
\end{equation}
Therefore, inflation has to be much more long-lasting (\mbox{$N_p\gsim 10^8$})
for this possibility to be realised, compared to the case of a light vector
field.

\section{Summary and conclusions}

In this paper we have studied in detail the inflationary buildup of an Abelian
vector boson condensate. Such a condensate, as we outlined in
Sec.~\ref{sec-stanis}, may be responsible for the quantitative predictions of a
cosmological model, which involves vector fields, such as statistical
anisotropy, either by mildly anisotropising the inflationary expansion
\cite{anisinf} or by involving directly the anisotropic vector field
perturbations in the curvature perturbation
\cite{Dimopoulos:2006ms,Dimopoulos:2008yv,yokosoda}.

In our treatment, we have mainly focused in the case of a vector field with
a time-varying kinetic function $f(t)$ and mass $m(t)$. This was partly
motivated by supergravity but it was also motivated by the peculiar type of
particle production of vector boson perturbations, which could be drastically
different from the case of a scalar field. We put emphasis on the possibility
that \mbox{$f\propto a^{-4}$} and \mbox{$m\propto a$}, which results in a flat
superhorizon spectrum of perturbations for both longitudinal and transverse
components, and may be an attractor if time-variation is due to the rolling
inflaton \cite{jacques}. The flat superhorizon spectrum of perturbations is
dominated by the longitudinal modes and, in contrast to the scalar field case,
its amplitude is decreasing with time even though it remains flat. As a result,
the condensate builds up onto a decreasing core as shown in Eq.~(\ref{Wtotal}).
Also, the condensate never equilibrates, albeit the vector field being massive,
in contrast to the well known Bunch-Davies result \cite{BD}
\mbox{$\langle\phi^2\rangle\sim H^4/m^2$}, for a massive but light ($0<m<H$)
scalar field. We have applied our findings to the vector curvaton mechanism as
an example, and showed that, if the condensate buildup is considered, we
obtain constraints on the total duration of inflation, as encoded in
Eq.~(\ref{Npbound}), if we want to generate observable statistical anisotropy.
This demonstrates the predictive power of this approach, compared to the
previous literature, which takes the value of the condensate as a free
parameter.

We also studied the buildup of an Abelian vector boson condensate in other
models of vector field
particle production and found some interesting results. For example,
we have looked into the time-varying $f$ and $m$ model when $f\propto a^2$, which
also produces scale invariant spectra for the vector field components. In this
case, we found that the condensate does equilibrate in a similar manner to the
light massive scalar field case, because the mass of the physical vector boson
is now constant. Another case we have looked into is the case of an Abelian
vector field non-minimally coupled to gravity through an $RA^2$ term, where we
found that the scale invariant case (coupling \mbox{$\gamma\approx 1/6$})
leads to a condensate buildup \mbox{$\langle W^2\rangle\sim H^3t$}, similar
to the massless scalar field case \cite{Linde:1982uu}. Finally, we looked also
into the case of an axial coupling and found that the vector condensate can be
exponentially amplified in the string axion inspired case when the spectrum
of the transverse vector field perturbations is flat and uneven.

Apart from the specific, model dependent results above, our work is the only
comprehensive study to date of the inflationary buildup of a vector boson
condensate and can be used as a blueprint by any future similar study
(see also Ref.~\cite{juan+kari}). We carry out our study by extending the methods of stochastic inflation (usually applied to scalar fields) to include vector fields. Owing to the different boundary conditions imposed on the various polarisation modes $w_\lambda$, we identify differences (with respect to the scalar field case) making necessary to modify the stochastic formalism to properly account for the evolution of the classical vector field $\mbox{\boldmath$W$}_c$. The bottom line of our method, developed in Sec.~\ref{sec-eee}, consists in introducing the conjugate momentum $\mbox{\boldmath$\Pi$}_\lambda$, to subsequently eliminate it in the equation for $\mbox{\boldmath$W$}_\lambda$ (Eq.~(\ref{eq82})) using the superhorizon behavior of the perturbation modes $w_\lambda$. Our method goes beyond the Hamiltonian description of stochastic inflation since we manage to obtain a \emph{single} first order equation for $\mbox{\boldmath$W$}_\lambda$ (Eq.~(\ref{eq32})) which, in turn, leads to a Fokker-Planck equation in the variable $\mbox{\boldmath$W$}_\lambda$ only. Finally, we remark that our procedure can be successfully applied to scalar fields with a non-negligible scale-dependence (i.e. the case of a heavy field) and also to phases of inflation away from the slow-roll regime.

All in all, we have investigated in detail the buildup of a vector boson field
condensate during inflation. We considered a multitude of Abelian vector field
models, where the conformal invariance of the field is appropriately broken, but
focused mostly onto the case of a time-varying kinetic function and mass. As an
example, we have applied our findings onto the vector curvaton mechanism and
obtained specific predictions about the duration and scale of inflation, which
were previously ignored when the magnitude of the condensate was taken as a
free parameter.

\section{Acknowledgements}
JCBS is supported by the Spanish Ministry of Science and Innovation through the research projects FIS2006-05895 and  Consolider EPI CSD2010-00064.
KD is supported (in part) by the Lancaster-Manchester-Sheffield Consortium for
Fundamental Physics under STFC grant ST/J000418/1.

\appendix

\section{\boldmath The case of $\nabla (A_t)_c$}
\label{appA}

Expanding $\delta A_\mu$ using creation/annihilation operators as in
Eq.~(\ref{eq4}), the temporal component $A_t$ is determined by [c.f.
Eq.~(\ref{eq16})]
\[
A_t=\int\frac{d^3k}{(2\pi)^3}\,\frac{k}{k^2+a^2M^2}\,\partial_t
\left\{f^{-1/2}a\left[\hat a_\parallel(k)w_\parallel\,e^{i\mbox{\scriptsize
\boldmath$k$}\cdot \mbox{\scriptsize
\boldmath$x$}}+\hat a_\parallel^\dagger(k)w_\parallel^*\,e^{-i\mbox{\scriptsize
\boldmath$k$}\cdot \mbox{\scriptsize
\boldmath$x$}}\right]\right\}\,.
\]
To compute $(\mbox{\boldmath $\nabla$}A_t)_c$ we multiply the above integrand
by $\theta(k_s-k)$ to extract the long wavelength part and utilize the
superhorizon limit of $w_\parallel$ in Eq.~(\ref{eq11}). Taking the gradient and
writing $\mbox{\boldmath$k$}=\mbox{\boldmath$e_\parallel$}\,k$ we arrive at
\[\label{eq25}
(\mbox{\boldmath $\nabla$}A_t)_c=
\int\frac{d^3k}{(2\pi)^3}\,\theta(k_s-k)\frac{k^2\partial_t
\left[f^{-1/2}a\left(
\mbox{\boldmath $e$}_\parallel
\,\hat a_\parallel(k)w_\parallel\,
e^{i\mbox{\scriptsize\boldmath$k$}\cdot \mbox{\scriptsize
\boldmath$x$}}+\mbox{\boldmath $e$}_\parallel^*
\,\hat a_\parallel^\dagger(k)w_\parallel^*\,e^{-i\mbox{\scriptsize
\boldmath$k$}\cdot \mbox{\scriptsize
\boldmath$x$}}\right)\right]}{k^2+a^2M^2}\,.
\]
Since the longitudinal modes dominate over the transverse ones [c.f.
Eqs.~(\ref{eq24}) and (\ref{eq11})], Eq.~(\ref{eq5}) indicates that the terms
$\mbox{\boldmath$\ddot{W}$}_c$ and $3H\mbox{\boldmath$\dot{W}$}_c$ are of order
$H^2\mbox{\boldmath$W$}_c\gg M^2\mbox{\boldmath$W$}_c$. Consequently, the term
$p(t)\mbox{\boldmath $\nabla$}(A_t)_c$ in Eq.~(\ref{eq12}) can be neglected
provided that
\[\label{eq27}
\left|p(t)(\mbox{\boldmath $\nabla$}A_t)_c\right|\ll
\left|M^2(t)\mbox{\boldmath$W$}_c\right|\,.
\]
Since we can approximate \mbox{\boldmath$W$}$_c$ as the superposition of
longitudinal modes only, i.e.
\begin{equation}\label{eq20}
\mbox{\boldmath $W$}_{\!c}(t, \mbox{\boldmath $x$})\simeq
\int\frac{d^3k}{(2\pi)^
3}\;\theta(k_s-k)\left[
\mbox{\boldmath $e$}_\parallel
\,\hat a_\parallel(\mbox{\boldmath $k$})
w_\parallel e^{i\mbox{\scriptsize\boldmath $kx$}}
+\mbox{\boldmath $e$}_\parallel^*
\,\hat a_\parallel^\dagger(\mbox{\boldmath $k$})
w_\parallel^* e^{-i\mbox{\scriptsize\boldmath $kx$}}\right]\,,
\end{equation}
to find out whether (\ref{eq27}) is satisfied it suffices to compare the square
brackets in Eqs.~(\ref{eq25}) and (\ref{eq20}). If we denote by
$w_\parallel^{(d)}$ the decaying, albeit dominant on superhorizon scales, part of
$w_\parallel$ (second term in Eq.~(\ref{eq13})), then we have
$f^{-1/2}aw_\parallel^{(d)}\propto a^3w_\parallel^{(d)}\simeq{\rm cte}$ for modes above
the coarse-graining scale . Consequently, only the growing part of $w_\parallel$
(first term in Eq.~(\ref{eq13})), which we denote by $w_\parallel^{(g)}$,
contributes to the gradient operator in Eq.~(\ref{eq25}). Operating the
integrand in Eq.~(\ref{eq25}) we obtain
\[
\left|\frac{\partial_t\left[f^{-1/2}a\left(\hat a_\parallel(k)w_\parallel^{(g)}+
\hat a_\parallel^\dagger(k)w_\parallel^{(g)*}\right)\right]}{1+(aM/k)^2}\,p(t)\right|=
\frac{24H^2\left(\hat a_\parallel(k)w_\parallel^{(g)}+\hat a_\parallel^\dagger(k)
w_\parallel^{(g)*}\right)}{1+(aM/k)^2}\,,
\]
where we used that $w_\parallel^{(g)}\simeq{\rm cte}$ on superhorizon scales.
Using now the expression of $w_\parallel^{(g)}$ that follows from
Eq.~(\ref{eq13}), the condition (\ref{eq27}) translates into
\[
(k/aH)^3\ll1+(aM/k)^2\,,
\]
which holds in the superhorizon regime. Therefore, we may neglect the gradient
of the temporal component $A_t$ to describe the evolution of
$\mbox{\boldmath$W$}_c$. Note that this is an expected result since for
sufficiently superhorizon scales ($r\gg r_c\gg1$) the equations of motion for
$w_{L,R}$ and $w_\parallel$ coincide [c.f. Eqs.(\ref{eq15}) and (\ref{eq8})].

\section{Direct computation}\label{appB}
The general solution to the non-homogeneous equation
\[
\dot W_{\lambda}+{\cal F}_\lambda(t) W_{\lambda}\simeq\xi_{W_\lambda}\,,
\]
can be easily obtained
\[\label{eq42}
W_\lambda(t)=\exp\left[-\int_0^t{\cal F}_\lambda (\tau)\,d\tau\right]W_\lambda(0)+
\int_0^t\exp\left[-\int_{\bar \tau}^t {\cal F}_\lambda(\tau)\,d\tau\right]\,
\xi_{W_\lambda}(\bar\tau)\,d\bar\tau\,.
\]
If \mbox{\boldmath$\xi$}$_{W_\lambda}$ is a white noise source, i.e.
$\langle\mbox{\boldmath$\xi$}_{W_\lambda}\rangle=0$, the ensemble average (over
independent representations of the stochastic source) is [c.f. Eq.~(\ref{eq41})]
\[
\langle W_\lambda(t)\rangle=\exp\left[-\int_0^t{\cal F}_\lambda (\tau)\,
d\tau\right]\,W_\lambda(0)\,.
\]
Using Eq.~(\ref{eq42}), the two-point function is [c.f. Eqs.~(\ref{eq49}) and
(\ref{eq47})]
\begin{eqnarray}
\langle W_\lambda^2\rangle&=&\langle W_\lambda\rangle^2+\int_0^t
\exp\left[\int_0^{\bar
\tau}{\cal F}_\lambda(\tau)\,d\tau\right]\int_0^{t}\exp\left[\int_0^{\hat \tau}
{\cal F}_\lambda(\tau)\,d\tau\right]\langle\xi_{W_\lambda}(\hat\tau)\,\xi_{W_\lambda}
(\bar \tau)\rangle\,d\hat \tau\,d\bar \tau\nonumber\\
&=&\langle W_\lambda\rangle^2+\int_0^t\exp\left[-2\int_{\bar\tau}^t
{\cal F}_\lambda(\tau)\,d\tau\right]\,D_\lambda(\bar\tau)\,d\bar\tau\,.
\end{eqnarray}

\end{document}